\documentclass[11pt]{article} 
\usepackage{a4}
\usepackage{amsmath}
\usepackage{graphicx}
\def\be{\begin{equation}}
\def\ee{\end{equation}}
\def\bea{\begin{eqnarray}}
\def\eea{\end{eqnarray}}

\def\C{{\rm\kern.24em
    \vrule width.02em height1.4ex depth-.05ex
    \kern-.26em C}}
\def\N{{\rm I\kern-.18em N}}
\def\R{{\rm I\kern-.21em R}}
\def\Z{{\rm\kern.26em
    \vrule width.02em height0.5ex depth 0ex
    \kern.04em
    \vrule width.02em height1.47ex depth-1ex
    \kern-.34em Z}}
\def\d{{\rm\kern.22em
    \vrule width.02em height1.0ex depth0ex
    \kern-.24em d}}

\renewcommand\slash[1]{\not \! #1}

\def\C{{\rm\kern.24em
    \vrule width.02em height1.4ex depth-.05ex
    \kern-.26em C}}
\def\N{{\rm I\kern-.18em N}}
\def\O{{\rm\kern.24em
    \vrule width.02em height1.45ex depth-.05ex
    \kern-.26em O}}
\def\P{{\rm I\kern-.25em P}}
\def\R{{\rm I\kern-.21em R}}
\def\Z{{\rm\kern.26em
    \vrule width.02em height0.5ex depth 0ex
    \kern.04em
    \vrule width.02em height1.47ex depth-1ex
    \kern-.34em Z}}

\renewcommand\slash[1]{\not \! #1}
\def\slashed{\slash}
\DeclareMathOperator{\Tr}{Tr}
\newcommand{\vc}{\mathbf}

\newdimen\picraise
\newcommand\picb[1]
{
  \setbox0=\hbox{\includegraphics{#1}}
  \picraise=-0.5\ht0
  \advance\picraise by 0.5\dp0
  \advance\picraise by 3pt     % correct for height of `=' above baseline
  \hbox{\raise\picraise \box0}
}

\newdimen\picraise
\newcommand\picresize[2]
{
  \setbox0=\hbox{\includegraphics[width=#2]{#1}}
  \picraise=-0.5\ht0
  \advance\picraise by 0.5\dp0
  \advance\picraise by 3pt     % correct for height of `=' above baseline
  \hbox{\raise\picraise \box0}
}

\newdimen\picraiseano
\newcommand\eqeps[2]
{
  \setbox0=\hbox{\includegraphics{#1}}
  \picraiseano=-0.5\ht0 
  \advance\picraiseano by 0.5\dp0
  \advance\picraiseano by -#2pt  % correct for height of pictures in inteq. 
  \hbox{\raise\picraiseano \box0}
}

\newdimen\picraiseano
\newcommand\eqepsres[3]
{
  \setbox0=\hbox{\includegraphics[width=#3]{#1}}
  \picraiseano=-0.5\ht0 
  \advance\picraiseano by 0.5\dp0
  \advance\picraiseano by -#2pt  % correct for height of pictures in inteq. 
  \hbox{\raise\picraiseano \box0}
}

\newdimen\picraise
\newcommand\picbox[1]
{
  \setbox0=\hbox{\input{#1}}
  \picraise=-0.5\ht0 
  \advance\picraise by 0.5\dp0
  \advance\picraise by 3pt     % correct for height of `=' above baseline
  \hbox{\raise\picraise \box0}
}

\newdimen\picraiset
\newcommand\picding[1]
{
  \setbox0=\hbox{\input{#1}}
  \picraiset=-0.5\ht0 
  \advance\picraiset by 0.5\dp0
  \advance\picraiset by -3pt  % correct for height of pictures in inteq. 
  \hbox{\raise\picraiset \box0}
}

\newdimen\picraisehallo

\newcommand\pichallo[2]
{
  \setbox0=\hbox{\input{#1}}
  \picraisehallo=-0.5\ht0 
  \advance\picraisehallo by 0.5\dp0
  \advance\picraisehallo by -#2pt  % correct for height of pictures in inteq. 
  \hbox{\raise\picraisehallo \box0}
}

\begin{document}

\begin{titlepage}
\begin{flushright}
HD-THEP-04-49\\
IFUM-792-FT\\
hep-ph/0501192
\\
\end{flushright}

\vfill

\begin{center}
\boldmath
{\LARGE{\bf The $C$-Odd Four-Gluon State in the}}
\\[.2cm]
{\LARGE{\bf Color Glass Condensate}}
\unboldmath
\end{center}
\vspace{1.2cm}
\begin{center}
{\bf \Large
Stefan Braunewell\,$^{a,b,1}$,  Carlo Ewerz\,$^{a,c,2}$
}
\end{center}
\vspace{.2cm}
\begin{center}
$^a$
{\sl
Institut f\"ur Theoretische Physik, Universit\"at Heidelberg\\
Philosophenweg 16, D-69120 Heidelberg, Germany}
\\[.5cm]
$^b$
{\sl
Institut f\"ur Theoretische Physik, Universit\"at Bremen\\
Otto-Hahn-Allee 1, D-28359 Bremen, Germany}
\\[.5cm]
$^c$
{\sl
Dipartimento di Fisica, Universit{\`a} di Milano and INFN, Sezione di Milano\\
Via Celoria 16, I-20133 Milano, Italy}
\end{center}                                                                 
\vfill

\begin{abstract}
\noindent
The study of the perturbative Odderon at high gluon densities 
in the color glass condensate 
requires to take into account states with more than three 
gluons. We formulate evolution equations for these 
states in which the number of gluons is not fixed during the evolution. 
We further determine the coupling of these Odderon states to the 
$\gamma \to \eta_c$ impact factor for arbitrary numbers of gluons. 
We find an exact solution of the evolution equation for the four-gluon 
Odderon state in terms of the three-gluon Odderon state. It is shown 
that a similar solution exists also for a different class of Odderon solutions 
which does not couple to the $\gamma \to \eta_c$ impact factor. 
We discuss the implications 
of our result from the perspective of constructing an effective 
field theory of reggeized gluons for the color glass condensate. 
\vfill
\end{abstract}
\vspace{5em}
\hrule width 5.cm
\vspace*{.5em}
{\small \noindent 
$^1$ email: braunewell@itp.uni-bremen.de \\
$^2$ email: Carlo.Ewerz@mi.infn.it 
}

\end{titlepage}

\section{Introduction}
\label{sec:intro}

In high energy collisions the partons inside hadrons form a very 
dense system which is often called the color glass condensate. 
At sufficiently high density the partons overlap such that recombination 
effects become important and tend to slow down the growth of 
the density with increasing energy. It is expected that eventually 
a saturation regime is reached. The theoretical challenge 
is to understand at which energies and how this takes place in detail. 
In order to answer this question one formulates evolution equations 
describing the behavior of the system with increasing energy. 
The color glass condensate has been studied in a number of different 
perturbative approaches which are applicable 
if the scattering process involves a hard scale. 
Prominent examples are the approach initiated by 
McLerran and Venugopalan 
\cite{McLerran:1993ni,McLerran:1993ka} (for a review 
see \cite{Iancu:2003xm}), 
the operator expansion of Wilson lines due to Balitsky 
\cite{Balitsky:1995ub}, or the 
color dipole picture of high energy scattering 
developed by Mueller  
\cite{Mueller:1993rr,Mueller:1994jq}. 
Most of these approaches lead to similar results in suitable 
approximations. They all reproduce for example the 
Balitsky-Kovchegov (BK) evolution equation which 
resums certain classes of multi-Pomeron exchanges  
\cite{Balitsky:1995ub,Balitsky:1998ya,Kovchegov:1999yj,Kovchegov:1999ua}, 
and which has become a key tool in investigating the 
color glass condensate and in particular potential 
saturation effects at high energies. 

The approach which we will use in the present paper is the 
generalized leading logarithmic approximation (GLLA) 
\cite{Bartels:1978fc}-\cite{Kwiecinski:1980wb}. 
It is an extension of the classical approach to high energy 
scattering in QCD based on the resummation of large 
logarithms that in leading logarithmic order (LLA) 
gives rise to the BFKL Pomeron \cite{Kuraev:fs,Balitsky:ic}. 
The BFKL Pomeron can be viewed as the exchange 
of two interacting reggeized gluons. In the GLLA one 
goes beyond that approximation by taking into account 
also exchanges of more than two gluons, and one collects 
all perturbative contributions that contain the maximally 
possible number of logarithms for a given number of gluons. 
There are two versions of the GLLA: in the first version 
the number of gluons exchanged in the $t$-channel of the 
scattering process remains constant, whereas in the 
second version the number of gluons is allowed to change  
during the $t$-channel evolution \cite{Bartels:unp}. 
The latter is often called extended GLLA, or EGLLA. 
It is the fluctuation of the number of gluons in the $t$-channel 
which in the resummation approach reflects the 
parton recombination effects characterizing the color glass condensate. 
Different aspects of the EGLLA have been studied 
in \cite{Bartels:1992ym}-\cite{Bartels:2004hb}. 
There it has been found that in the Pomeron channel, 
i.\,e.\ for the exchange of even $C$-parity, 
the EGLLA gives rise to a picture of an effective 
field theory in which states consisting 
of even numbers of gluons are coupled to each other via 
effective vertices. So far the vertices from two to four 
\cite{Bartels:1994jj} and from two to six gluons 
\cite{Bartels:1999aw} have been calculated explicitly. 
From the two-to-four gluon vertex one obtains the 
perturbative triple Pomeron vertex \cite{Lotter:1996vk}, 
and in a similar way the one-to-three Pomeron 
vertex is obtained from the two-to-six gluon vertex 
\cite{Ewerz:2003an}. 
Recently it has been shown that the triple Pomeron 
vertex obtained in this way gives rise 
to the BK equation when the so-called M\"obius 
representation is used for the Pomerons 
\cite{Bartels:2004ef}. 
As discussed in that reference the EGLLA not only 
reproduces the BK equation but also makes it possible 
to compute subleading corrections to the BK equation 
which appear difficult to access in other approaches to 
the color glass condensate. 

An advantage of the EGLLA is that the remarkable 
property of conformal invariance in two-dimensional 
impact parameter space 
\cite{Lipatov:1985uk,Bartels:1995kf,Ewerz:2001uq} 
and the phenomenon of gluon reggeization 
\cite{Lipatov:zz} 
become particularly transparent in this approach. 
Especially the reggeization of the gluon will turn out 
to be crucial in our present investigation. 
The gluons exchanged in high 
energy scattering are reggeized, that is they are composite objects 
consisting themselves of reggeized gluons in a 
selfconsistent way. A manifestation of this phenomenon 
is that whenever two $t$-channel gluons are at the same 
point in impact parameter space they behave like a single 
gluon. When this occurs explicitly in a scattering amplitude 
we can hence see how the composite 
gluon is formed out of the two gluons. The same picture 
can be extended to more than two gluons, and one can 
systematically resolve higher Fock states of the reggeized gluon 
in the framework of the EGLLA, see \cite{Ewerz:2001fb}. 

So far the color glass condensate has been studied almost 
exclusively in the Pomeron channel, that is for the exchange 
of vacuum quantum numbers (in particular for positive 
$C$-parity) which is relevant to total cross sections. 
In the present paper we make a first step of 
a systematic study of the Odderon in the EGLLA, that is 
for the channel in which negative charge parity quantum 
number is exchanged. In lowest order in the GLLA the Odderon is 
an exchange of three interacting gluons in the $t$-channel 
in a symmetric color state, and is described by the 
Bartels-Kwieci{\'n}ski-Prasza{\l}owicz (BKP) equation 
\cite{Bartels:1980pe,Kwiecinski:1980wb}. Two types of explicit 
solutions to the BKP equation have been found in 
\cite{Janik:1998xj} and \cite{Bartels:1999yt}, respectively. 
The latter solution has also been found in the dipole picture of 
high energy scattering in \cite{Kovchegov:2003dm} 
where also some potential effects of the color glass condensate on the 
Odderon are discussed. 
For a review of the theory and phenomenology of 
the Odderon see \cite{Ewerz:2003xi}. 

Our motivation for studying also the Odderon in the 
EGLLA is twofold. Firstly, states with more than three 
gluons in the $t$-channel might be important phenomenologically. 
One could for example expect that Pomeron loops can 
affect the Odderon intercept. Recall that the intercept of the 
perturbative Odderon is close to one (and exactly one 
for the solution of \cite{Bartels:1999yt}). Hence potential corrections to 
the intercept due to Pomerons would be particularly significant. 
Another interesting effect is the interplay of Pomeron and Odderon 
exchanges in high energy scattering. Some possible effects of that 
interplay have recently been discussed in \cite{Brodsky:2004qa}. 
A detailed study certainly requires a concise 
knowledge of the splitting of an Odderon into an Odderon 
and a Pomeron, as it can be obtained in the EGLLA. 
The calculation of that vertex is among the 
goals of the investigation that we initiate in the present paper. 
We expect that this vertex will also make it possible to obtain 
the large-$N_c$ limit of the EGLLA and thus give rise to 
an equation for the Odderon channel which is the analog 
of the BK equation for the Pomeron channel. This equation 
should then be of a similar form as the one suggested in 
the context of saturation in the dipole picture for the Odderon 
in \cite{Kovchegov:2003dm}. 

An equally important motivation for studying the Odderon in the 
EGLLA is to gain a better understanding of the effective theory 
of the color glass condensate and of its properties. Important aspects 
in this respect would for instance be an investigation of the crossing 
properties of the vertices in the effective field theory and their 
interpretation in view of conformal field theories. 
In addition, studying Odderon states will hopefully 
also make it possible to compute the gluon vertices of 
the effective field theory -- which so far are known only for 
color singlet channels -- in arbitrary color states. 

In a first step we consider in the present paper the case of up to four 
gluons. We give explicitly the coupled evolution equations for the 
three- and four-gluon states. 
As an initial condition for the evolution we choose the 
coupling of the Odderon to the $\gamma \to \eta_c$ impact factor. 
This impact factor can be computed in perturbation 
theory due to the large mass of the charm quark. It 
is also of phenomenological relevance, see for example 
\cite{Czyzewski:1996bv}-\cite{Braunewell:2004pf}. 
We will show that the impact factor exhibits reggeization. 
That fact will make it possible to find an explicit solution of 
the evolution equation for the four-gluon state when it is coupled 
to the $\gamma \to \eta_c$ impact factor. Due to that impact 
factor, the Odderon solution found in \cite{Bartels:1999yt} 
is projected out. We will show that an analogous result 
can be obtained for the solution of \cite{Janik:1998xj} by 
considering the case of a baryonic impact factor. 
As we will discuss in the conclusions, our result lays the 
foundations for a future investigation of the five-gluon 
Odderon state in which we expect new elements of the 
effective field theory to occur. 

We should mention that 
Odderon states with more than three gluons have already 
been discussed in the GLLA, but only with the number of gluons 
being kept fixed during the evolution. There a remarkable 
simplification takes place in the large-$N_c$ limit in which 
these $n$-gluon states are equivalent to an 
integrable XXX Heisenberg model 
\cite{Lipatov:1993preprint}-\cite{Faddeev:1994zg}. 
It should be pointed out, however, that for $n>3$ the 
$n$-gluon states obtained in the large-$N_c$ 
limit are not the leading ones at high energies. Moreover, 
they cannot be coupled directly to the phenomenologically 
interesting impact factors, as will become clear also in our 
discussion below. We emphasize that in the present paper 
we do not fix the number of gluons in the $t$-channel 
evolution, and also do not take the large-$N_c$ limit. 

The paper is organized as follows. 
In section \ref{sec:eveq} we formulate the evolution 
equations for the $n$-gluon Odderon states in the EGLLA. 
We then compute the $\gamma \eta_c$-Odderon impact factor 
with an arbitrary number $n$ of gluons ($n \ge 3$) and give the 
result explicitly for the cases of four and five gluons. The result 
is expressed in terms of the impact 
factor with only three gluons. Based on that observation we 
are then able to find an analytic solution of the four-gluon Odderon 
state in the EGLLA and discuss its properties in section 
\ref{sec:solution}. In section \ref{sec:jw} we generalize our 
result to other impact factors and hence to all known 
types of Odderon solutions. Our results are 
summarized in section \ref{sec:summary}. 
In the course of our calculation we also find some useful results 
for the $C$-even channel which we present in an appendix. 

\section{Coupled evolution equations for the Odderon in EGLLA}
\label{sec:eveq}

We start with the well-known BKP equation for a system of 
three interacting gluons exchanged in the $t$-channel of a 
scattering process \cite{Bartels:1980pe,Kwiecinski:1980wb}. 
In the case of the Odderon the three gluons 
are in a state that is odd under $C$-parity. 
We consider the Odderon amplitude coupled to the $\gamma \to \eta_c$ 
impact factor, which hence constitutes the initial condition for the 
evolution of the Odderon state in rapidity. We define $F_3$ as the 
amputated three-gluon Odderon amplitude with discontinuities taken 
in the squared energies obtained from the four-momenta 
of the photon and the first gluon, and of the photon and the first and 
second gluon, respectively. At high energies, the dynamics effectively 
reduces to the transverse plane of the reaction, see section 
\ref{simpfacrep} below. The amplitude 
$F_3$ accordingly depends on the transverse momenta of the 
three gluons, and it obviously carries color labels for the gluons. 
In addition, $F_3$ depends on the momenta of the photon and the 
$\eta_c$-meson, but we will suppress this dependence in our notation. 
It is convenient to change from the squared center-of-mass energy $s$ 
to complex angular momentum $\omega$ via a Sommerfeld-Watson 
transformation. 

The BKP equation in transverse momentum space then reads 
\begin{equation}
\label{inteqf3}
  \left(\omega - \sum_{i=1}^3 \beta(\vc k_i) \right) F_3^{a_1 a_2 a_3} =
F_{(3;0)}^{a_1 a_2 a_3} 
+ \sum K^{\{b\} \rightarrow \{a\}}_{2\rightarrow 2} \otimes F_3^{b_1b_2b_3}
\,,
\end{equation}
with the gluon trajectory function $\beta$ describing virtual corrections 
to the $t$-channel gluons, 
\begin{equation}
\label{traject}
  \beta(\vc k^2) = - \frac{N_c}{2} g^2  \int \frac{d^2\vc l}{(2 \pi)^3} 
          \frac{\vc k^2}{\vc l^2 (\vc l -\vc k)^2} 
\,, 
\end{equation}
where the bold-face characters denote two-dimensional transverse momenta. 
The BKP equation can be viewed as a Schr\"odinger type equation with 
$\omega$ being the energy-like variable. The conjugate time-like 
variable is rapidity $Y=\log(s/s_0)$, where $s_0$ is 
a fixed hadronic energy scale. The kernel 
$K^{\{b\} \rightarrow \{a\}}_{2\rightarrow 2}$ describes the pairwise 
interaction of the gluons. It is up to a color factor identical to the part 
of the integral kernel of the BFKL equation which describes real gluon 
production. We will give an explicit 
representation of the kernel as well as an explanation of the convolution 
symbol further below. 
In the case of three gluons one can separate the color and the 
momentum part, 
\begin{equation}
\label{splitf3}
  F_3^{a_1 a_2 a_3} (\vc k_1,\vc k_2,\vc k_3) = 
d_{a_1a_2a_3} F_3 (\vc k_1,\vc k_2,\vc k_3) 
\,, 
\end{equation}
with the symmetric structure constant $d_{a_1a_2a_3}$ of the 
$\mbox{SU}(3)$ color group. 
We will sometimes suppress the momentum arguments of the function $F_3$. 

The inhomogeneous term $F_{(3;0)}$ in the BKP equation (\ref{inteqf3}) 
is given by the impact factor of the transition $\gamma \to \eta_c$.
We refer the reader to \cite{Czyzewski:1996bv} for the explicit formula for the 
impact factor with three gluons. Its explicit form will not be needed for 
our discussion. We only note here that the impact 
factor $F_{(3;0)}$ is symmetric in the three gluon momenta and 
in their color labels, and that it vanishes due to gauge invariance 
when one of the gluon momenta vanishes. 
It should be pointed out that our particular choice 
of the $\gamma \to \eta_c$ impact factor singles out a particular 
solution to the BKP equation, namely the Bartels-Lipatov-Vacca 
(BLV) solution \cite{Bartels:1999yt}. That solution is a superposition 
of states depending only on two transverse coordinates. In contrast 
to that situation, the other known class of solutions 
of the BKP equation requires the three gluons to be at 
different positions in transverse space and vanishes when two of the 
three gluon positions coincide. 
(The Janik-Wosiek solution found in \protect\cite{Janik:1998xj} belongs to 
this class.) 
In the $\gamma \to \eta_c$ impact factor in leading order the photon splits 
into a quark-antiquark pair to which the three gluons couple, 
and which then recombines into an $\eta_c$ meson. The intermediate 
state with only two quarks provides only two points in transverse 
space to which the gluons couple and hence singles out the BLV solution. 
We will first concentrate on that solution and will come back to the 
other class of solutions in section \ref{sec:jw}. 

The symmetry of the impact factor mentioned above immediately 
allows us to find a crucial property of the full three-gluon 
amplitude without even using the particular form of the BLV solution, 
namely the separate invariance of the 
amplitude under exchange of the momenta and of the color indices.
The latter observation is trivial because of the color tensor $d_{a_1a_2a_3}$ 
which the Odderon keeps throughout its evolution. 
The invariance under momentum-exchange can be inferred from the BKP 
equation (\ref{inteqf3}) as the quark loop has this property 
and the application of the integral kernel 
respects the symmetry because the sum over all pairwise interactions is 
performed. Formally constructing the solution by iterating
the integral equation then leads to the conclusion that this property transfers
to the full three-gluon Odderon amplitude. Of course, this symmetry property 
is also reflected in the explicit formula of the BLV solution, see \cite{Bartels:1999yt}. 
Later we will need only a slightly weaker property of the three-gluon 
Odderon state, namely its symmetry under simultaneous exchange of
momentum and color labels which is a trivial consequence of the separate 
symmetries in color and momentum space. 
In a similar way one can start from the fact that the 
impact factor vanishes for vanishing gluon momenta and derive that 
also the full amplitude $F_3$ vanishes if one of the three gluon momenta 
vanishes. 

We now want to write the BKP equation in a diagrammatic form in order 
to make the generalization to higher gluon numbers more transparent, 
\begin{equation}
\label{inteqf3diag}
\left( \omega - \sum_{i=1}^3\beta(\vc k_i)\right) 
\eqeps{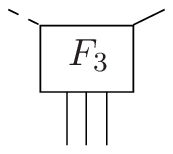}{-2} =
\eqeps{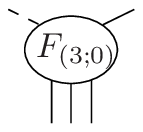}{-1} 
+ \sum \eqeps{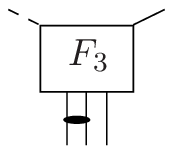}{-2} \,.
\end{equation}
In this notation, the diagrams representing the impact factor and the full
Odderon amplitude include the color tensors and the contributions from the 
external particles, but are amputated, i.e.\ the `outgoing' gluon propagators
are cut off. The kernel that acts on the amplitude also includes 
color tensors which we give explicitly below. 
The sum in the last term extends over all pairwise interactions of the gluons.

It is now straightforward to apply the EGLLA to the $C$-odd channel 
following the same procedure which has been developed in 
\cite{Bartels:1980pe,Bartels:unp} for the $C$-even (Pomeron) channel. 
The lowest possible contribution in the $C$-odd channel is the 
Odderon with three gluons satisfying the BKP equation, see above. 
In the EGLLA, we have to take into account in addition all exchanges 
with more than three gluons, and have to allow for number-changing 
transitions during the evolution. This is most conveniently 
implemented in the form of integral equations that generalize the 
BKP equation. (For a detailed discussion of the corresponding 
integral equations in the $C$-even channel see \cite{Bartels:1999aw}.) 
In these integral equations the number-changing transitions 
are due to integral kernels $K_{2\to m}$ which have been derived 
in \cite{Bartels:unp}. Note that as a result of the approximation 
scheme of the EGLLA in these kernels only two gluons 
interact with each other to produce more $t$-channel gluons. 
This is similar to the BKP equation (\ref{inteqf3}), where in each 
contribution to the last term only two gluons 
undergo an interaction via the BFKL kernel. 
In the amplitude for an $n$-gluon state one then has to take 
into account all possible contributions in which a state with 
$l$ gluons ($l<n$) undergoes a transition to an $n$-gluon state via 
a kernel $K_{2\to m}$ with $m=n-l+2$, and we have to include 
all possible $l$ starting from the lowest possible amplitude that 
has $l=3$.  
For us it is important that the derivation \cite{Bartels:unp} 
of the kernels $K_{2\to m}$ for the interaction of two gluons 
does not require any assumptions about the other gluons, 
in particular it does not require a specific $C$-parity of the 
whole $n$-gluon state. 
We can therefore use exactly the same kernels to formulate the integral 
equations for $n$-gluon amplitudes also in the $C$-odd channel. 

In this way one obtains the following integral equation for the $C$-odd 
four-gluon amplitude $F_4$ in the EGLLA, 
\begin{eqnarray}
\label{inteqf4}
  \left( \omega - \sum_{i=1}^4 \beta(\vc k_i) \right) 
  F_4^{a_1a_2a_3a_4} &=& F_{(4;0)}^{a_1a_2a_3a_4} 
  + \sum K^{\{b\} \rightarrow \{a\}}_{2\rightarrow 3} \otimes F_3^{b_1b_2b_3} 
  \nonumber \\
  &&+ \sum K^{\{b\} \rightarrow \{a\}}_{2\rightarrow 2} 
  \otimes F_4^{b_1b_2b_3b_4}   \,,
\end{eqnarray}
% $\alpha(\vc k^2) = 1 + \beta(\vc k^2)$ 
which in diagrammatic notation reads:
\begin{equation}
\label{inteqf4diag}
\left( \omega - \sum_{i=1}^4\beta(\vc k_i)\right) 
\eqeps{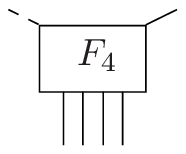}{-2} =
\eqeps{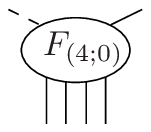}{-3} 
+ \sum \eqeps{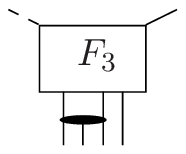}{-2}
+ \sum \eqeps{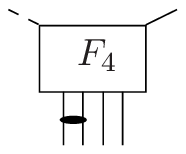}{-2}.
\end{equation}
One can easily construct also the equations for the higher $n$-gluon 
amplitudes $F_n$, again in complete analogy to the case of even 
$C$-parity. For $n=5$, for example, the equation reads 
\begin{eqnarray}
\label{inteqf5}
  \left( \omega - \sum_{i=1}^5 \beta(\vc k_i) \right) 
  F_5^{a_1a_2a_3a_4a_5} &=& F_{(5;0)}^{a_1a_2a_3a_4a_5} 
  + \sum K^{\{b\} \rightarrow \{a\}}_{2\rightarrow 4} \otimes F_3^{b_1b_2b_3} 
  \nonumber \\
  &&+ \sum K^{\{b\} \rightarrow \{a\}}_{2\rightarrow 3} 
  \otimes F_4^{b_1b_2b_3b_4}   
  \nonumber \\
&& + \sum K^{\{b\} \rightarrow \{a\}}_{2\rightarrow 2} 
\otimes F_5^{b_1b_2b_3b_4b_5}   
\,.
\end{eqnarray}
In this way one obtains the full (infinite) hierarchy of 
coupled integral equations of the EGLLA for the $C$-odd channel. 

To complete the account of the integral equations for the 
$C$-odd channel in EGLLA, we
have to give the exact expression for the interaction kernels. 
As the $2\to m$ kernels can be written in a general form, we can do this
in one step for both kernels that appear up to the four-gluon equation.
Let the two gluons that enter the kernel from above carry 
the transverse momenta $\vc l_1$ and $\vc l_2$
and the color labels $b_1$ and $b_2$, and let the $m$ outgoing gluons carry
the momenta $\vc k_1,\dots,\vc k_m$ and the color labels $a_1, \dots, a_m$.
The integral kernels for the transition of two to $m$ gluons 
then read \cite{Bartels:unp} 
(the generalization to arbitrary indices being trivial)
\begin{equation}\label{2tomkernel}
\begin{split}
  K_{2\to m}^{\{b\}\to\{a\}} (\vc l_1, \vc l_2; \vc k_1, \dots, \vc k_m)
   = & \,\, g^m \, f_{b_1 a_1 k_1} f_{k_1 a_2 k_2} \dots f_{k_{m-1} a_m b_2}\\
  & \left[ (\vc k_1 + \dots + \vc k_m)^2 
    - \frac{\vc l_2^2(\vc k_1+\dots+\vc k_{m-1})^2}{(\vc k_m-\vc l_2)^2}
  \right. \\
   &- \left. \frac{\vc l_1^2(\vc k_2+\dots+\vc k_m)^2}{(\vc k_1-\vc l_1)^2}
    + \frac{\vc l_1^2\vc l_2^2 
      (\vc k_2+\dots+\vc k_{m-1})^2}{(\vc k_1-\vc l_1)^2(\vc k_m-\vc l_2)^2}
  \right]\,. 
\end{split}
\end{equation}
For the $2\to 2$ kernel ($m=2$) the last term in the 
brackets is defined to be zero 
so that one gets the BFKL kernel without the virtual corrections.
Finally, the convolution symbol $\otimes$ implies 
a factor $[(2\pi)^3 \vc l_1^2 \vc l_2^2]^{-1}$ followed by 
an integration $\int d^2 \vc l_1$ over the loop momentum. 
Clearly, there is a color
factor $\delta^{a_i b_j}$ for every gluon that does not participate in the
kernel interaction. 

As in the case of $C$-even exchanges, the integral equations of 
the EGLLA as described above apply to cut amplitudes. The four-gluon 
amplitude $F_4$, for example, is a result of taking three discontinuities 
with respect to the energy variables obtained from the photon and 
the first $r$ gluons, $r=1,2,3$. It should be noted that, 
as a result of these discontinuities, the amplitude $F_4$ is not fully 
symmetric under the exchange of two gluons. The full Bose symmetry 
of the $n$-gluon states is only restored in the full amplitude reconstructed 
via dispersion relations. Instead, the cut amplitudes for which the integral 
equations are formulated obey a particular set of Ward-type 
identities found in \cite{Ewerz:2001fb} for the $C$-even channel. 
As we will discuss below, our result for $F_4$ satisfies these 
identities. 

Finally, a remark is in order concerning the calculation of 
scattering amplitudes from the amplitudes $F_n$ of our 
integral equations. In order to calculate the amplitude for the 
quasidiffractive 
process $\gamma\gamma\to \eta_c \eta_c$ for three $t$-channel 
gluons, one would have to fold the solution of the BKP 
equation (\ref{inteqf3}) simply with the lower impact factor after 
attaching simple gluon propagators (without interaction) to it 
as is illustrated in figure \ref{famplgen} below (see also 
\cite{Braunewell:2004pf}). 
However, for the higher gluon number Odderon equations such a simple  
reconstruction of the full scattering amplitude must fail, because the 
integral equation also includes the transition from three to four 
$t$-channel gluons but not the other way around, which would lead to 
an asymmetric treatment of the upper and lower impact factor. 
Simply taking the square of the result is also not 
possible because it would double count some of the contributions. 
In fact, the reconstruction of the physical amplitude for the process 
$\gamma\gamma\to \eta_c \eta_c$ from the solution $F_n$ of 
the integral equations for general $n$ is a nontrivial task and 
requires a very careful use of dispersion relations in order to undo 
the cuts applied to the amplitude. A detailed discussion of this problem 
is beyond the scope of the present paper. 
When considering other scattering processes like for example
photon-proton scattering, however, the coupling of the four-gluon 
system directly to the scattering partner is easier and does not 
lead to the problems mentioned above.

\boldmath
\section{The $\gamma \eta_c$-Odderon impact factor}
\unboldmath
\label{sec:impfac}

We now turn to the calculation of the $\gamma \to \eta_c$ impact 
factor with an arbitrary number $n >3$  of gluons coupled to it 
in GLLA. Due to the quantum numbers of the photon and of the 
$\eta_c$ meson the $n$ gluons are always in a $C$-odd state. 

It will be useful to recall first in a more general setting 
the factorization of scattering processes at high energies which 
gives rise to impact factors. 
After discussing the $\gamma \to \eta_c$ impact factor 
with three gluons we will then consider more gluons. 
For that the crucial point 
is to study the step from $n$ to $n+1$ gluons, that is the 
effect of attaching one additional gluon to 
the impact factor. We show that the corresponding 
diagrams with $n+1$ 
gluons can be expressed in terms of diagrams with 
$n$ gluons in a particular way. That step has been considered 
more or less explicitly in many studies before (see for example 
\cite{Braun:1995hh,Bartels:1999aw}), 
but we find it useful to discuss it here 
in more detail in a notation suitable for our purposes. 
Using that result and taking into account the color 
algebra we are then able to find 
a general formula for our impact factor with $n$ gluons 
in terms of the three-gluon impact factor. 

\subsection{High energy factorization and impact factors}
\label{simpfacrep}

Let us consider the scattering amplitude for a perturbatively 
calculable process at high energies in the GLLA. 
To be specific, we choose the quasidiffractive process 
$\gamma \gamma \to \eta_c \eta_c$, but the discussion 
also holds for other processes. 
In the following derivation of the impact factor representation 
for the scattering amplitude we follow \cite{LF71} 
(where the analysis was conducted for the QED case, i.e.\ photon exchange) 
and \cite{Ginzburg:1985tp} (where the two-gluon QCD case is treated).

At high energies the scattering process is dominated by gluon 
exchange, the lowest order contribution being three-gluon exchange 
for the process under consideration. 
\begin{figure}[ht]
 \begin{center}
\includegraphics[width=7cm]{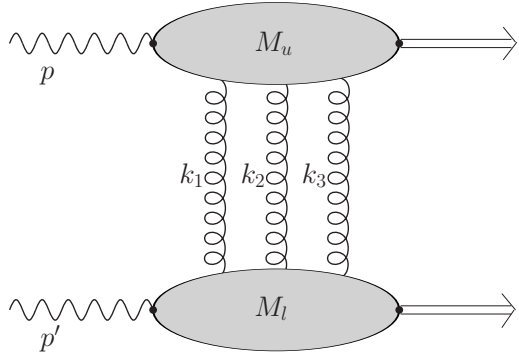}
\caption{Diagrammatic representation of the scattering amplitude for a simple
three-gluon exchange process.}
\label{famplitude}
\end{center}
\end{figure}
Let us first concentrate on the simple case in which the three 
gluons do not interact with each other, as illustrated in figure 
\ref{famplitude}. 
We can write the corresponding amplitude as 
\begin{equation}\label{amplitude}
\mathcal{M} = \frac{i}{3!}\int \frac{d^4 k_1}{(2 \pi)^4} \frac{d^4
 k_2}{(2 \pi)^4} M^{\mu\nu\rho}_u \frac{(-{\rm i} 
g_{\mu \alpha})}{k_1^2}
\frac{(-ig_{\nu \beta})}{k_2^2} \frac{(-ig_{\rho \gamma})}{k_3^2} 
M^{\alpha\beta\gamma}_l\,,
\end{equation}
where $3!$ is a combinatorial factor and 
$M^{\mu\nu\rho}_u$ and $M^{\mu\nu\rho}_u$ stand for 
the upper and lower shaded parts of the figure, respectively. 
These vertex functions also 
include the color factors and the contribution of the external particles. 

Explicitly writing out the arguments of these functions 
for the case of incident photons, we have
\begin{align}
M^{\mu\nu\rho}_u & \equiv M^{\mu\nu\rho}_u(p,\epsilon,k_1,k_2,k_3) \\
M^{\alpha\beta\gamma}_l & \equiv M^{\alpha\beta\gamma}_l
(p',\epsilon',-k_1,-k_2,-k_3)\,.
\end{align}
The minus signs in front of the gluon momenta in the term 
$M^{\alpha\beta\gamma}_l$ arise because an incoming momentum in the
upper vertex is treated as outgoing in the lower vertex and vice versa.
The polarization vectors $\epsilon$ and $\epsilon'$ of the incident photons 
will not be relevant in our discussion and will therefore not be written 
explicitly below. The three gluon momenta are related to the 
momentum transfer via $q=k_1+k_2+k_3$, and $q^2=t$. 

Now we perform a Sudakov decomposition of the momenta by splitting
each four-momentum into its components parallel to the 
two light-like four-vectors
$p$ and $p'$ that have antiparallel three-momentum directions, and the 
remaining transverse part. For example, the vector $k_1$ is decomposed into
\begin{equation}\label{Sudakov} 
k_1=\alpha_1 p + \beta_1 p' + k_{1T}\,.
\end{equation}
In high-energy collisions we can assume that the invariant masses
are negligible compared to $s$. Hence 
$s = (p + p')^2 \approx 2 p \cdot p'$ and we can use the incident 
photon momenta for the decomposition also in the case of virtual 
photons. We will always denote by $\alpha$ the Sudakov component 
belonging to the upper incident momentum $p$, and by $\beta$ the 
one belonging to $p'$. The integration measure can then be rewritten as
\begin{equation}
  d^4 k_i = p \cdot p' \, d\alpha_i \, d\beta_i \, d^2 
\mathbf k_{i}
 = \frac{1}{2}s \,  d\alpha_i \, d\beta_i \, d^2 \mathbf k_{i}, \qquad i=1,2
\,.
\end{equation}
Furthermore, the transferred
momenta are predominantly transverse (as was shown for the QED case
in \cite{LF71}), 
hence the denominator of a gluon propagator can be simplified to
\begin{equation}
k_i^2 = s \alpha_i \beta_i + k_{iT}^2 \approx - \mathbf k_i^2\,.
\end{equation}
Also the metric tensor $g^{\mu\alpha}$ can be divided into longitudinal and
transverse parts, 
\begin{equation}
  g_{\mu\alpha}=\frac{2}{s}(p_\mu p'_\alpha + p_\alpha
 p'_\mu) +
 g^T_{\mu\alpha}\,.
\end{equation}

After the convolution with the vertex functions $M^{\mu\nu\rho}_u$ and 
$M^{\alpha\beta\gamma}_l$, the large contributions arise from the terms 
in which the momentum of the lower incident particle is contracted with
the upper vertex, and vice versa. Phrased differently, one finds that 
the longitudinal gluon polarizations dominate in the high energy limit. 
We can therefore substitute
\begin{equation}
  g_{\mu\alpha} g_{\nu\beta} g_{\rho\gamma} \to \frac{8}{s^3}
 p'_\mu p'_\nu p'_\rho p_\alpha  p_\beta
 p_\gamma\,,
\end{equation}
and obtain the amplitude in impact factor representation,
\begin{equation}\label{Amplf}
\mathcal M = \frac{s}{3} \int \frac{d^2\mathbf k_1}{(2 \pi)^2} 
\frac{d^2\mathbf k_2} {(2 \pi)^2} \Phi_u 
\frac{1}{\mathbf k_1^2}
\frac{1}{\mathbf k_2^2} \frac{1}{\mathbf k_3^2}  \Phi_l\,,
\end{equation}
with the impact factors being
\begin{eqnarray}
\Phi_u &=& \int \frac{d\beta_1}{2\pi}\frac{d\beta_2}{2\pi}
M_u^{\mu\nu\rho}\frac{p'_\mu p'_\nu p'_\rho}{s}, \label{upperimpfac}\\
 \Phi_l &=& \int \frac{d\alpha_1} {2\pi}\frac{d\alpha_2} {2\pi}
 M_l^{\alpha\beta\gamma}\frac{p_\alpha p_\beta p_\gamma}{s}\,.
\end{eqnarray}
The integrals can be disentangled here because the 
$\alpha$-parameters of the gluons can be neglected in $M_u^{\mu\nu\rho}$ 
(they are small compared to the $\alpha$-parameters of the quark lines), 
and likewise for the $\beta$-parameters in the lower vertex. 

\begin{figure}[ht]
\begin{center}
\includegraphics[width=7cm]{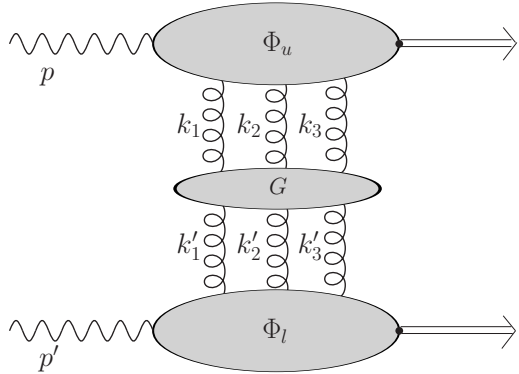}
\caption{Impact factor representation of a scattering process with 
$t$-channel gluon interaction.}
\label{famplgen}
\end{center}
\end{figure}

The above considerations are easily generalized to more sophisticated 
exchanges in which the $t$-channel gluons interact with each other 
as is shown in figure \ref{famplgen}. 
Also here the dominant contributions come from the longitudinally 
polarized gluons and the resulting impact factors are exactly 
the same as above. The resulting amplitude can be written 
symbolically as 
\begin{equation}\label{convolution}
\mathcal M = \frac{s}{3}\langle \Phi_u |\mathbf G | \Phi_l \rangle.
\end{equation}
In that general case the matrix element symbol stands for the integration
over all undetermined momenta of the scattering amplitude -- that is, 
$\mathbf k_1$, $\mathbf k_2$, $\mathbf k'_1$ and $\mathbf k'_2$ (the third
gluon's momentum is fixed by the total momentum transfer). It also
includes a factor $(2 \pi)^{-4}$ as in the simple case discussed above. 
{\bf G} denotes the Green function of the three-gluon exchange in 
transverse space. In this picture the simple three-gluon exchange 
process (fig.\ \ref{famplitude}) is reproduced by setting
\begin{equation}
  \mathbf G = \frac{\delta^{2}(\mathbf k'_1-\mathbf k_1)
\delta^{2}(\mathbf k'_2-\mathbf k_2)}
{\mathbf k_1^2 \mathbf k_2^2 \mathbf k_3^2}\,.
\end{equation}

It is also straightforward to generalize the derivation of the impact 
factors to exchanges with an arbitrary number of gluons in the 
$t$-channel. From the formulae above one can easily read off 
the factors and integrations which have to be associated with each 
additional gluon. 

\subsection{Color algebra}
\label{sec:color}

Let us now collect some basics of color algebra for the gauge 
group SU($N_c$).  Its Lie algebra has $N_c^2-1$ 
generators $t^a$ that satisfy the algebra
\begin{equation}
  \big[t^a,t^b\big]=i f_{abc} t^c 
\end{equation}
with the antisymmetric structure constants $f_{abc}$. The anti-commutator of two
generators defines the symmetric structure constants $d_{abc}$, 
\begin{equation}
  \big\{t^a,t^b\big\}=\frac{1}{N_c} \delta^{ab} + d_{abc} t^c\,.
\end{equation}
Note that we write the structure constants with lower indices. 

Sometimes we will use the so-called `bird track' notation
to illustrate color tensor contractions diagrammatically. 
The antisymmetric structure constants are drawn as a solid black circles 
and the color indices are written counterclockwise:
\begin{equation}
  f_{abc} = \picresize{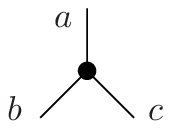}{1.3cm}\,.
\end{equation}
Then for example the nontrivial part of the color tensor of the $2 \to 3$ kernel 
in (\ref{2tomkernel}) is given by
\begin{equation}
  f_{b_1 a_1 k_1} f_{k_1 a_2 k_2} f_{k_2 a_3 b_2} = 
\picb{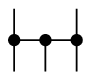}\,.
\end{equation}
The color tensors for the integral kernels are drawn in such a way 
that the color indices $b_i$, which are to be contracted with the color 
labels of the amplitudes in the integral equations, 
are at the top of the diagram.

It is convenient to define the tensors 
\begin{align}
  d^{a_1 a_2 \dots a_n}& = \Tr(t^{a_1} t^{a_2} \dots t^{a_n}) +
\Tr(t^{a_n} t^{a_{n-1}} \dots t^{a_1}),\\
\label{defftensor}
  f^{a_1 a_2 \dots a_n}& =\frac{1}{i} [\Tr(t^{a_1} t^{a_2} \dots t^{a_n}) - 
\Tr(t^{a_n} t^{a_{n-1}} \dots t^{a_1})].
\end{align}
Note that these definitions do not coincide with the structure constants
for $n=3$. In fact, the structure constants are given in terms of these 
tensors by 
\begin{align}
  d_{a_1 a_2 a_3}& = \frac{1}{2} d^{a_1 a_2 a_3},\\
  f_{a_1 a_2 a_3}& = \frac{1}{2} f^{a_1 a_2 a_3}.
\end{align}

\subsection{The Odderon quark loop with three gluons}
\label{impfactwiththree}

Next we want to review some important properties of the 
$\gamma \to \eta_c$ impact factor with three gluons 
attached to it which was first calculated in \cite{Czyzewski:1996bv}. 
Three is the lowest possible number of gluons 
in the $t$-channel for which that impact factor exists. 

\begin{figure}[ht]
 \begin{center}
\includegraphics[width=6cm]{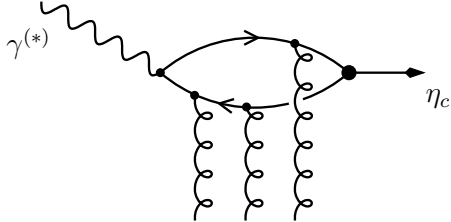}
\caption{One of the diagrams contributing to the $\gamma \to\eta_c$ 
impact factor with three gluons.}
\label{famplitudediag}
\end{center}
\end{figure}

The $\gamma \to \eta_c$ impact factor for three gluons consists of eight
diagrams corresponding to the $2^3$ choices for the way in which 
the gluons couple to the the quark or antiquark. (Recall that due to the 
cuts applied to the amplitude the gluons do not cross each other.) 
One of these diagrams is shown in figure \ref{famplitudediag}. 
Most of the details of the analytic expression for the impact factor 
will not be relevant for our considerations. A key property is 
that the Dirac structure of the coupling of the quark-antiquark 
pair to the $\eta_c$ meson is simply $\gamma^5$. 
The coefficient of the $\gamma^5$ factor can be related to 
the radiative two-photon width of the $\eta_c$, see 
\cite{Czyzewski:1996bv}. Further, it turns out that only 
transverse photon polarizations give a nonvanishing 
contribution to the impact factor. In the following we will often speak 
of the quark loop to which the Odderon is coupled. Unless otherwise 
stated this will always assume that there is a $\gamma_\mu$ matrix 
at the photon vertex and a $\gamma^5$ matrix at the meson vertex. 

When calculating one particular diagram (out of the eight possible diagrams) 
one encounters five quark propagators of the form
\begin{equation}
  \frac{i(\slashed k + m)}{k^2 - m^2}\,,
\end{equation}
four additional $\gamma$ matrices (three gluon vertices and the photon vertex)
and a $\gamma^5$ 
coming from the $\eta_c$ vertex. The gluon vertex $\gamma$ matrices get
contracted with the light-like vector $p'$ from the lower impact factor 
(see section \ref{simpfacrep}). 
Using the fact that for light-like vectors $p$
the equation $\slashed p \slashed p =0$ holds, one easily
reduces the expression in the $\gamma$ trace to a sum of terms that consist 
of maximally five $\gamma$'s by permuting 
the gluon vertex $\gamma$ matrices next to each other.

When performing the trace for the case of massless quarks, 
one clearly gets zero because of the odd number of $\gamma$ matrices, 
and hence we need to take into account mass effects. 
That allows us to choose a mass
term instead of a momentum term with a $\gamma$ matrix (in the numerator of the
quark propagator $\slashed {k} + m$). In fact, only the case
in which of the five propagator terms one mass and four momenta are combined
gives a non-vanishing contribution. Let us stress this point: when calculating
the impact factor, exactly one propagator term must lend its mass, the others
their momenta with $\gamma$ matrices. Clearly, we will get a sum over the
different possibilities of which propagator lends its mass, but for our  
purposes it will suffice to keep these considerations on a qualitative level.

Another important property is the behavior of an impact factor diagram under
the exchange of its quark and antiquark lines. We can reassign the quark loop
momenta in such a way that we get almost the same expression as before, 
apart from a minus sign
for all propagator momenta in the $\gamma$ trace. For three gluons this means
an overall relative {\it plus} sign because one propagator must give a mass
instead of a momentum (as we have explained above) and the sign of the mass is 
not affected by the exchange. 
The set of all possible impact factor diagrams can be grouped 
into pairs in which the two diagrams are just related to each other 
by such an interchange of the quark and antiquark lines. 
Doing so, one reduces the number of 
diagrams that have to be calculated by two. 
This holds for the momentum and $\gamma$ part only, but
the interchange of the quark lines naturally also reverses the order of the
$SU(N_c)$ generators. 

For three gluons we finally get four diagrams in this way, each coming together with
two different traces of color matrices. Adding these two contributions one gets
\begin{equation}
\Tr (t^{a_3} t^{a_2} t^{a_1}) + \Tr (t^{a_1} t^{a_2} t^{a_3})
=d^{a_1 a_2 a_3} \,.
\end{equation}
The full impact factor is then just the sum over the four different momentum
diagrams together with the color tensor $d^{a_1 a_2 a_3}$:
\begin{equation}\label{f30}
\eqeps{glf32}{-3} = d^{a_1 a_2 a_3}\left(
\picb{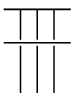}+
\picb{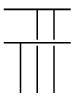}+
\picb{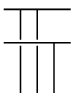}+
\picb{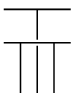}\right)\equiv d^{a_1 a_2 a_3} F_{(3;0)}\,.
\end{equation}

The schematic notation introduced here stands symbolically for the impact 
factor without color factors. The diagrams only show which gluons couple to
the quark (upper line) and which to the antiquark (lower line) -- the overall 
color tensor $d^{a_1 a_2 a_3}$ is extracted but the photon
and $\eta_c$ vertices are implied.
To obtain the final result, one needs to calculate these different diagrams.

As already said, 
we will not need the explicit form of the three-gluon impact factor. 
But there is one property that will be very important for our
considerations, namely 
the symmetry under exchange of its momentum arguments.
It does not hold for every contributing momentum diagram separately, but 
can only be seen after performing the sum over the different diagrams.

\subsection{Mechanism of gluon number reduction}
\label{sec:mechanism}

Now we want to explain how an $n+1$ gluon diagram can be reduced 
to an $n$ gluon diagram. 
Clearly we can always find at least two gluons that are 
attached to the same quark line. Hence we can choose two which are separated
only by one quark propagator. The additional factors that enter -- compared to
the $n$-gluon diagram -- are a gluon vertex contracted with the incident
momentum of the lower impact factor $p'$ and the additional quark 
propagator between the two neighboring gluon vertices. The propagator
again has a sum $\slashed k +m$ in its numerator but as the mass term is
sandwiched between two factors of $\slashed p'$ its contribution vanishes.
Therefore, the addition of another gluon increases the number of $\gamma$ 
matrices by two. 
To reduce the number of $\gamma$ matrices in the trace,
we use the same mechanism that we have already explained in the beginning
of the previous section. Thereby we can
always reduce the trace to one involving four $\gamma$ matrices, 
the $\gamma^5$ from the meson vertex, and one quark mass. 
In the following we want to have a closer look at this reduction 
mechanism and at the factors which occur. For that it will be 
important that we work with cut amplitudes, which amounts 
to putting the cut quark lines on-shell. 

First we choose two gluons that are attached to the same quark line at 
neighboring vertices. Writing it in Feynman diagram notation, 
this part of the quark loop can be represented as
\begin{equation}\label{2to1}
\eqeps{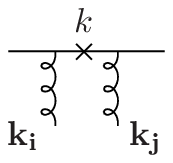}{-6}.
\end{equation}
Let $k$ be the momentum of the quark line that lies
between these two vertices. We will call the gluon further down along the
line `new' gluon, because we understand it as an additional gluon as compared
to the $n$-gluon amplitude. According to the Cutkosky rules for calculating
the discontinuity of the diagram, the `new' quark propagator term is 
replaced according to 
\begin{equation}
  \frac{i ({\slashed k}+m)}{k^2 -m^2} \to i ({\slashed k}+m) 
[-2 \pi i \delta(k^2-m^2)] \,,
\end{equation}
because a quark line between two neighboring gluons is always put on shell due
to the cuts.

Let $j$ be the index of the new gluon.
The new vertex gets contracted with the momentum of the lower incident particle
$p'$ to give $i g t^{a_j} {\slashed p}'$. 
As was explained in section \ref{simpfacrep}, for every gluon one also gets an
integration over its Sudakov component parallel to $p'$,
i.e.\ over $\beta_j$. Putting it all together, one finds as additional factors 
apart from the new color trace:
\begin{equation}
\begin{split}
  \dots \int \frac{d \beta_j}{2 \pi} i g \Tr (\dots i ({\slashed k}+m) 
  \slashed p' \dots ) [-2 \pi i \delta(k^2-m^2)] \\
= \dots i g \int d \beta_j
  \Tr (\dots ({\slashed k}+m) \slashed p' \dots ) \, \delta(k^2-m^2)\,.
\end{split}
\end{equation}

The neighboring gluon with momentum $\vc k_i$ in (\ref{2to1}) also gives a 
${\slashed p}'$ in the $\gamma$-trace, and we can permute it with the
quark line numerator and get a factor $2 \, p'\cdot k$.
By using the Sudakov decomposition for the quark line, 
$k=\alpha p + \beta p' + k_T$, this can be absorbed by the $\delta$-function, 
and so we get
\begin{equation}
  i g \int d \beta_j \Tr (\dots \slashed p' \dots ) \,
\delta\left(\beta - \frac{{\bf k}^2+m^2}{\alpha s}\right)\,.
\end{equation}

We now shift the integration variables 
\begin{align}
  \beta_i \to \ & \beta_i'=\beta_i + \beta_j,\\
  \beta_j \to \ & \beta_j.
\end{align}
It is clear that all the $\beta$ parameters in the quark loop apart from the 
ones coming from the new quark propagator can be expressed in terms of
$\beta_i'$ and the $\beta$ parameters of the other gluons and of the quark loop
momentum only, which means that the only occurrence of $\beta_j$ is in the new
propagator.\footnote{If one denotes the $\beta$ parameter of the
incoming quark momentum at the left of (\ref{2to1}) by $\tilde \beta$, 
one easily finds $\beta=\tilde \beta + \beta_i' -\beta_j$, so one can see that
$\beta_j$ in fact enters in the argument of the $\delta$-function.}
Now the integration over
$\beta_j$ can be performed to cancel the $\delta$-function. In pictorial 
notation, we are left with the following identity:
\begin{equation}\label{diagreduction}
\eqeps{2to1gluon1}{-6} =
i g \eqeps{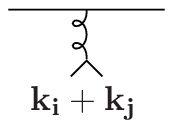}{1}\,.
\end{equation}

So far we have discussed only the momentum part of one particular diagram
that contributes to the Odderon impact factor with $n+1$ gluons.
We also have to consider the relative sign between two diagrams that are linked
by interchange of the quark and antiquark lines. Compared to the $n$-gluon
case we have included another quark propagator and a $\gamma$ matrix coming
from the new gluon vertex. 
Independently of the number of
attached gluons, always one mass term in the $\gamma$ trace 
has to be chosen, so every new propagator gives a new relative minus sign since 
the momentum term changes sign under the interchange of the fermion lines. 
(Note that this is due to the fact that we are calculating a closed fermion loop in 
which the direction of the momentum flow is reversed due to the interchange 
of the fermion lines.) 
Remembering that for 3 gluons we found a
relative {\it plus} sign, we easily see that the relative sign between diagrams
with interchanged quark and antiquark lines is $(-1)^{n+1}$. In other 
words, the momentum part of a diagram with an odd number of gluons is 
symmetric under the interchange of quark and antiquark, 
whereas an even number of gluons
leads to an antisymmetric momentum part. Thus, we
get $d$-type color tensors for odd numbers of attached gluons, 
$f$-type tensors for even
numbers (with a factor of $i$ to cancel the $1/i$ in the definition 
(\ref{defftensor}) of the $f$ tensors).

\subsection{The Odderon quark loop with four gluons}
\label{sec:loopfour}

We are now ready to express the $n+1$ gluon quark loop for the Odderon
in terms of the one with $n$ gluons. 
The main calculation becomes clear in the step from three to four gluons.
It is then easy to proceed to higher gluon numbers.
The four-gluon impact factor consists of two color structures, each 
with four pairs of momentum diagrams.
The terms in which all four gluons are attached to the
same fermion line, for example, give the following structure (all traces are
performed in the direction opposite to the quark lines):
\begin{equation}
  \Tr(t^{a_4} t^{a_3} t^{a_2} t^{a_1} - t^{a_1} t^{a_2} t^{a_3} t^{a_4} )
\picb{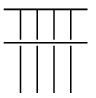}=-i f^{a_1 a_2 a_3 a_4} \picb{f40oooo}\,,
\end{equation}
where we have reversed the order of the color indices of the $f$ tensor,
which gives an additional minus sign.
For the complete four-gluon quark loop one finds the 
following diagrammatic representation:
\begin{equation}
\begin{split}
\label{f40}
\eqeps{glf42}{-3} = \,  - i & f^{a_1 a_2 a_3 a_4}\left(
\picb{f40oooo}+
\picb{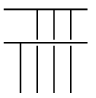}+
\picb{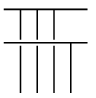}+
\picb{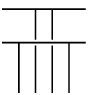}\right)\\
 \,  - i & f^{a_3 a_1 a_2 a_4}\left(
\picb{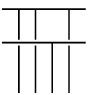}+
\picb{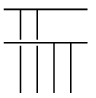}+
\picb{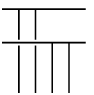}+
\picb{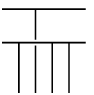}\right)\\ \equiv
  -i & f^{a_1 a_2 a_3 a_4} F_{(4;0)}^{(1)}
 - \, i f^{a_3 a_1 a_2 a_4} F_{(4;0)}^{(2)} \,.
\end{split}
\end{equation}

Clearly, there are many possible ways of reducing the single momentum 
diagrams to three-gluon diagrams along the lines of the previous section. 
But it is our aim to express 
it in terms of the full three-gluon impact factor $F_{(3;0)}$. 
Motivated by the structure of the three-gluon $\gamma \to \gamma$ 
impact factor in the $C$-even channel \cite{Bartels:1992ym} and 
by the general property of reggeization in the $n$-gluon amplitudes 
in that channel \cite{Bartels:1999aw,Ewerz:2001fb} we now make an 
ansatz and show that it in fact gives the full impact factor. 
The following construction leads to the correct 
result for the four-gluon Odderon impact factor:  
\begin{equation}
\label{f40ansatz}
\eqeps{glf42}{-3} = \frac{g}{2} \left[
\eqeps{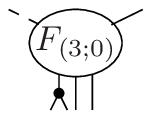}{-1}+
\eqeps{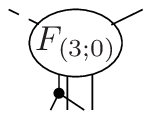}{-1}+
\eqeps{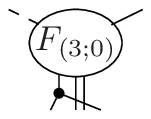}{-1}+
\eqeps{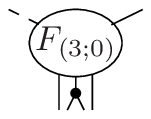}{-1}+
\eqeps{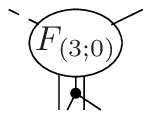}{-1}+
\eqeps{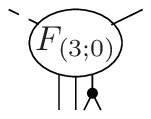}{-1}\right] \,.
\end{equation}
This diagrammatic notation symbolizes the contraction of the color tensor
of the impact factor $F_{(3;0)}$ with an $f_{abc}$ tensor. 
In addition, one of the
arguments of the amplitude $F_{(3;0)}$ is actually the sum of two gluons'
momenta. We will further abbreviate this expression containing 
sums of two arguments and contractions 
with a color tensor $f$ in the following way:
\begin{equation}\label{F40ansatzsum}
\eqeps{glf42}{-3} = 
\frac{g}{2}\sum_{i,j \in \{1,\ldots,4\} \atop i < j} 
\left[\eqeps{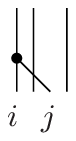}{-2} \right]
\star F_{(3;0)}^{b_1 b_2 b_3} (i j) \,,
\end{equation}
where $i$ and $j$ denote the position of the two `merging' gluons in the 
four-gluon system and the same indices appear in the color tensor and in the
sum of the momenta. The star symbolizes the contraction of the color tensor
(drawn in bird track notation) with the tensor of the amplitude
$F_{(3;0)}^{b_1 b_2 b_3}$.
The arguments of the function $F_{(3;0)}$ are very much simplified here.
Clearly, the momenta of all gluons enter as arguments of the function,
but the main point on which we focus 
in this notation is that the momenta of the two merged gluons enter only with
their sum. To shorten the notation, we simply write down the indices that
correspond to the merged gluons. That notation means that, for example, 
\be
 F_{(3;0)}^{b_1 b_2 b_3} (12) = 
 F_{(3;0)}^{b_1 b_2 b_3} (12,3,4) = 
F_{(3;0)}^{b_1 b_2 b_3} ({\bf k}_1+{\bf k}_2,{\bf k}_3,{\bf k}_4) \,.
\ee
A string of indices stands for the sum of the corresponding momenta. 
As we have mentioned before, the three-gluon impact factor is symmetric 
under the exchange of its momentum arguments, so we do not need to worry
about the position at which the merged gluons' sum enters as long as we choose
it consistently in all the specific diagrams shown in (\ref{f30}). 
With this abbreviation of the momentum arguments, expression (\ref{f40ansatz})
has the following explicit form,
\begin{equation}
\begin{split}
%\begin{align}
\label{f40ansatzexplicit}
F_{(4;0)}^{a_1 a_2 a_3 a_4}=\frac{g}{2} \Big[ & f_{a_1 a_2 k}
  F_{(3;0)}^{k a_3 a_4}(12) + f_{a_1 a_3 k} F_{(3;0)}^{k a_2 a_4}(13)
  + f_{a_1 a_4 k} F_{(3;0)}^{k a_2 a_3}(14)\\
+ & f_{a_2 a_3 k} F_{(3;0)}^{a_1 k a_4}(23) + f_{a_2 a_4 k}
  F_{(3;0)}^{a_1 k a_3}(24) + f_{a_3 a_4 k} F_{(3;0)}^{a_1 a_2
    k}(34)\Big] \,. 
%\end{align}
\end{split}
\end{equation}
%where we have used the $\delta$-symbol to write the amplitudes
%on the right hand side with color indices corresponding to the outgoing
%gluons.

To prove (\ref{f40ansatz}) one writes out all 24 diagrams of three-gluon 
quark loops according to (\ref{f30}). The momentum part of these
diagrams is easy to handle. A pair of gluons that couple to a quark line together
is just expanded to their respective positions in the four-gluon amplitude
as shown in (\ref{diagreduction}), corresponding to a factor $\frac{-i}{g}$.
Two gluons that enter the antiquark line give an additional minus sign when 
expanded. 

One might think that a problem arises when a gluon is
sandwiched by two merged gluons as is the case for example in the second
diagram on the right hand side of (\ref{f40ansatz}), 
because we have discussed only the case
when two neighboring gluons at the same quark line reduce to one. 
Recall, however, that we can
reduce {\it all} gluons that are attached to the same quark line to one.
Only the sum of the momenta of the gluons then enters the amplitude. Therefore,
we can use the mechanism described above also for the general case of any two 
gluons attached to the same quark line.

The color part is slightly more involved, because the $d^{b_1 b_2 b_3}$ 
tensor of the
three-gluon quark loop gets contracted with the color tensor which is given
in (\ref{F40ansatzsum}) in bird track notation. For example, if the first
two gluons are merged, this tensor is $f_{b_1 a_1 a_2} \delta^{b_2 a_3}
\delta^{b_3 a_4}$. In order to treat the tensors of this type we use the identity 
\begin{equation}
  f_{abk} d^{kcd} = f^{abcd}-f^{bacd} \,.
\end{equation}
Applying it to all the color contractions and using the cyclic invariance of the
$f$ tensors and the antisymmetry under reversal of all color indices, 
one ends up with three different tensors: 
$f^{a_1 a_2 a_3 a_4}$, $f^{a_3 a_1 a_2 a_4}$, and $f^{a_1 a_3 a_2 a_4}$.
If one now collects all diagrams that come with these respective 
color tensors, some contributions cancel (because the interchange of 
quark and antiquark line induces a sign change), 
others get a factor of 2. The diagrams coming with the color 
tensor $f^{a_1 a_3 a_2 a_4}$ cancel completely.
We are then exactly left with the expression in (\ref{f40}), which 
confirms that our ansatz (\ref{f40ansatz}) was indeed correct. 

In section \ref{sec:solution} we will promote the ansatz of 
(\ref{f40ansatz}) to an ansatz for the full four-gluon amplitude 
$F_4$. However, for the future project of solving the integral 
equations in the $C$-odd channel with more than four gluons 
a different representation for the impact factor might also be 
useful. We therefore also want to give that form which does not 
make use of the explicit color tensor $d^{b_1b_2b_3}$ 
of the three-gluon impact factor. This is helpful
because the quark loops for higher gluon numbers can be easily
written in terms of the momentum part of the three gluon expression.

In order to obtain that representation we go back one
step and write down the expression for the four-gluon quark loop in terms
of only the momentum part $F_{(3;0)}$ of the three-gluon amplitudes:
\begin{equation}
\begin{split}
  F_{(4;0)}^{a_1 a_2 a_3 a_4}=\frac{g}{2} \{ & f^{a_1 a_2 a_3 a_4} 
[F_{(3;0)}(12)-F_{(3;0)}(14)+F_{(3;0)}(23)+F_{(3;0)}(34)]\\ 
+& f^{a_3 a_1 a_2 a_4} [F_{(3;0)}(12)-F_{(3;0)}(13)+F_{(3;0)}(24)
-F_{(3;0)}(34)] \} \,. 
\end{split}
\end{equation}
Here we can identify the momentum parts corresponding to the two color
tensors:
\begin{align}
\label{f401}
F_{(4;0)}^{(1)} & = \frac{ig}{2}  [F_{(3;0)}(12)-F_{(3;0)}(14)+F_{(3;0)}(23)
+F_{(3;0)}(34)]  \,,\\
F_{(4;0)}^{(2)} & = \frac{ig}{2}[F_{(3;0)}(12)-F_{(3;0)}(13)+F_{(3;0)}(24)
-F_{(3;0)}(34)] \,. \label{f402}
\end{align}
By directly looking at (\ref{f40}) an even simpler representation of 
$F_{(4;0)}^{(1)}$ can be found:
\begin{equation}\label{f401simpler}
  F_{(4;0)}^{(1)}=ig F_{(3;0)}(23) \,.
\end{equation}
and the four-gluon quark loop is expressed as
\begin{equation}
\begin{split}\label{f40fromf30}
  F_{(4;0)}^{a_1 a_2 a_3 a_4} = g & f^{a_1 a_2 a_3 a_4} F_{(3;0)}(23)\\
+ & \frac{g}{2}  f^{a_3 a_1 a_2 a_4} [F_{(3;0)}(12)-F_{(3;0)}(13)+F_{(3;0)}(24)
-F_{(3;0)}(34)] \} \,. 
\end{split}
\end{equation}

In summary, we have derived two different (but equivalent) representations of the
four-gluon quark loop in terms of the three-gluon quark loop: in the first
representation (\ref{f40ansatzexplicit}) 
the three-gluon quark loop with its color structure is used, 
and that will be very convenient for the construction of the four-gluon Odderon
amplitude in section \ref{sec:solution}. 
The second representation (\ref{f40fromf30})
uses only the momentum part of the three-gluon expression and is useful for 
the investigation of the Odderon integral equations for more than four gluons. 

\subsection{Generalization to an arbitrary number of gluons}
\label{sec:generaln}

We now consider the general case of a quark loop with $n$ attached 
gluons, $n>3$. 
It consists of $2^n$ different diagrams. As could already
be seen in the examples of three and four gluons in equations (\ref{f30})
and (\ref{f40}), four diagrams
always lead to the same color structure, because switching the first or last
gluon from the quark to the antiquark or vice versa 
clearly does not affect the color trace.
We then get $2^{n-2}$ different color traces for the $n$-gluon
quark loop. These diagrams differ in the relative attachment 
(to the quark or the antiquark) of the `inner' $n-2$ gluons. 
In the previous sections
we also grouped these $2^{n-2}$ different diagrams in pairs because the
contributions of two diagrams in which quark and antiquark are interchanged
differ only by a relative sign. For the purpose of the present section it will 
be more convenient to collect these pairs at the end. 

In detail the combinatorics of the inner gluons is as follows. 
If all inner gluons couple to the quark line, one has the color trace
$\Tr(t^{a_n} t^{a_{n-1}} \ldots t^{a_1})$. Then there are $n-2$ 
different diagrams in which one inner gluon couples to the antiquark, 
whereas all others couple to the quark, and so on. For $k$ inner gluons
attached to the antiquark and $n-2-k$ attached to the quark line,
we get $\binom{n-2}{k}$ different color traces. To obtain the correct
color trace one simply has to write (in descending order of the gluon indices)
the color matrices 
that correspond to the `quark' gluons in front of the matrices of the 
`antiquark' gluons (in ascending order). 
Summing all these contributions we get the complete set of
contributions. 

Now we have to take into account the pairing of diagrams with interchanged
quark and antiquark lines. For example, the diagram in which only the first
of the inner gluons couples to the antiquark line gives (apart from the 
relative sign and the reversed order of the color generators) the same result 
as the diagram in which all but the first inner gluon 
couple to the antiquark. We thus get $2^{n-3}$ color tensors of the type
$d$ or $f$ together with
four momentum diagrams each. We choose the convention to write down those 
diagrams of each pair where the first inner gluon is attached to the quark line, 
as we did also in the case of the diagrams of $F_{(4;0)}^{(2)}$, see (\ref{f40}). 

These momentum diagrams can now easily be 
expressed in terms of the two parts $F_{(4;0)}^{(1)}$ and $F_{(4;0)}^{(2)}$ 
of the four-gluon amplitude, see (\ref{f401}) and (\ref{f402}). 
If all inner gluons are attached to the same line, the resulting expression 
reads 
\begin{align}
 -if^{a_1 \dots a_n} (ig)^{n-3} & F_{(3;0)}(2\dots n-1) \qquad 
 \mbox{for even } n \,, \\
 d^{a_1 \dots a_n} (ig)^{n-3} & F_{(3;0)}(2\dots n-1) \qquad 
 \mbox{for odd } n \,,
\end{align}
where we have used (\ref{f401simpler}) to express it in terms of $F_{(3;0)}$.
We have also reversed the order of the color indices so that the gluons that 
are attached to the quark line are written in ascending order. This introduces
a minus sign for all $f$ tensors.

If at least one of the inner gluons couples to the antiquark, the structure naturally is
more complicated. We will explain the construction of the general form, 
but also give the example of five gluons along the way.

First one has to identify all $2^{n-3}-1$ additional combinations of how the
inner gluons can couple to the quark or antiquark lines. For five gluons these 
are 
\begin{equation}\label{f50step1}
    d^{a_4 a_3 a_1 a_2 a_5} \, \picb{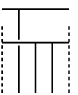}\,,  \quad
    d^{a_3 a_1 a_2 a_4 a_5} \, \picb{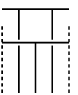}\,,  \quad
    d^{a_4 a_1 a_2 a_3 a_5} \, \picb{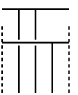}\,.
\end{equation}
We only show the coupling of the {\it inner} gluons here, because the 
first and the last gluon do not affect the color structure as was explained
above. Symbolically, we therefore draw the outer gluons as dashed lines and 
do not specify to which line they couple.

Let $i_1, \dots, i_k$ now be the indices of the inner gluons that are attached
to the antiquark line and $j_1, \dots, j_{n-2-k}$ the indices of the inner
gluons attached to the quark line. As explained before, the merging of two 
gluons at the antiquark line introduces an additional minus sign. Now we can 
write down the complete expression for one particular color structure,
\begin{equation}
  \begin{split}   \label{ngluoncomplete} 
     -i (-1)^{k-1} f^{a_{i_k} \dots a_{i_1} a_1 a_{j_1} \dots
      a_{j_{n-2-k}} a_n} (ig)^{n-4} 
& F_{(4;0)}^{(2)} (1,j_1 \dots j_{n-2-k}, i_1 \dots i_k,n) 
\quad \mbox{for even } n \,, \\
(-1)^{k-1} d^{a_{i_k} \dots a_{i_1} a_1 a_{j_1} \dots
      a_{j_{n-2-k}} a_n} (ig)^{n-4} 
& F_{(4;0)}^{(2)} (1,j_1 \dots j_{n-2-k}, i_1 \dots i_k,n) 
\quad \mbox{for odd } n \,.
\end{split}
\end{equation}
In order to avoid confusion here we have reinstated all four 
arguments of the amplitude $F_{(4;0)}^{(2)}$, and again 
the collection of indices in one argument
stands for the sum of the respective gluons' momenta. 
Adding all contributions, one arrives at the expression
for the general $n$-gluon quark loop in terms of three- and 
four-gluon quark loop amplitudes. 

As an example of this construction, 
the explicit form for the full impact factor for five gluons is 
\begin{equation}
\begin{split}
  F_{(5;0)}= (ig)^2 & d^{a_1 a_2 a_3 a_4 a_5} F_{(3;0)}(1,234,5) -
   ig d^{a_4 a_3 a_1 a_2 a_5} F_{(4;0)}^{(2)}(1,2,34,5)\\+
   ig & d^{a_3 a_1 a_2 a_4 a_5} F_{(4;0)}^{(2)}(1,24,3,5)+ 
  ig d^{a_4 a_1 a_2 a_3 a_5} F_{(4;0)}^{(2)}(1,23,4,5) \,.
\end{split}
\end{equation}

Finally, one can use (\ref{f402}) to express all the parts in
terms of the three-gluon amplitude $F_{(3;0)}$ only. 
As we have chosen the diagrams in such a way that the second
gluon is attached to the quark line, the sum of quark line gluons enters the
amplitude $F_{(4;0)}$ as the second argument and the sum of antiquark line 
gluons enters as the third argument. We thus arrive at the
expression for the different color parts of the $n$-gluon quark loop in terms
of the three-gluon quark loop $F_{(3;0)}$. For even $n$ it reads:
\begin{equation}
  \begin{split}
     -i \frac{(-1)^{k-1}}{2} (ig)^{n-3} & f^{a_{i_k} \dots a_{i_1} a_1 a_{j_1} \dots
      a_{j_{n-2-k}} a_n} \\
    [& F_{(3;0)} (1 j_1 \dots j_{n-2-k},i_1 \dots i_k, n)
    - F_{(3;0)} (1 i_1 \dots i_k, j_1 \dots j_{n-2-k}, n)\\
    + & F_{(3;0)} (1, j_1 \dots j_{n-2-k} n, i_1 \dots i_k)
    -F_{(3;0)} (1, j_1 \dots j_{n-2-k}, i_1 \dots i_k n)] \,.
  \end{split}
\end{equation}
For odd $n$, once again the $f$ color tensor has to be replaced by a $d$ 
tensor and the factor $-i$ has to be dropped. 
To get the full impact factor, all $2^{n-3}$ contributions have to be added.

Again, we want to give the final expression for the five-gluon case to 
illustrate our results:
\begin{equation}
\begin{split}
  F_{(5;0)}= -g^2 & d^{a_1 a_2 a_3 a_4 a_5} F_{(3;0)}(1,234,5) \\
  + \frac{g^2}{2} & d^{a_4 a_3 a_1 a_2 a_5} 
  [F_{(3;0)}(12,34,5)- F_{(3;0)}(134,2,5)+F_{(3;0)}(1,25,34)\\
& \hspace*{1.8cm}
-F_{(3;0)}(1,2,345)] \\
  - \frac{g^2}{2} & d^{a_3 a_1 a_2 a_4 a_5} 
  [F_{(3;0)}(124,3,5)- F_{(3;0)}(13,24,5)+F_{(3;0)}(1,245,3)\\
& \hspace*{1.8cm}
-F_{(3;0)}(1,24,35)] \\
  - \frac{g^2}{2} & d^{a_4 a_1 a_2 a_3 a_5} 
  [F_{(3;0)}(123,4,5)- F_{(3;0)}(14,23,5)+F_{(3;0)}(1,235,4)\\
& \hspace*{1.8cm}
-F_{(3;0)}(1,23,45)] \,. \\
\end{split}
\end{equation}

In summary, we have computed the $\gamma \to \eta_c$ impact 
factor for an arbitrary number of gluons. We have shown how 
it can be expressed in terms of the three-gluon impact factor. 
For each $n$ the impact factor is a superposition of terms in 
which in the momentum part a subset of the gluons behaves 
like a single gluon in the three-gluon impact factor, that is only 
the sum of their momenta enters. 

In appendix \ref{app:pomloop} we discuss how the construction 
of the $n$-gluon quark loop can be performed for the 
$\gamma \to \gamma$ impact factor, that is in the $C$-even 
sector, in a completely analogous way. 

\section{Solution for the four-gluon Odderon amplitude}
\label{sec:solution}

In \cite{Bartels:1993ih, Bartels:1994jj} it was observed that the 
three-gluon amplitude $D_3$ in the EGLLA is a superposition 
of two gluon amplitudes $D_2$, 
\begin{equation}
  D_3^{a_1 a_2 a_3}(\vc k_1, \vc k_2, \vc k_3)= 
\frac{g}{2} f_{a_1 a_2 a_3}
  [D_2(12,3)-D_2(13,2)+D_2(1,23)] \,.
\end{equation}
This result required that the impact factor $D_{(3;0)}$ describing 
the $\gamma^* \to \gamma^*$ transition with three $t$-channel 
gluons could be expressed in terms of the two-gluon impact 
factor $D_{(2;0)}$, see (\ref{d30}). A similar observation was made 
in a part of the five-gluon state in the $C$-even sector, namely in the 
part containing the two-to-four gluon transition vertex, see 
\cite{Bartels:1999aw}. In both cases the solution is given as 
a superposition of lower amplitudes in each of which a pair of gluons 
combines into a single gluon. In \cite{Ewerz:2001fb} this 
reggeization of the gluon was discussed also for more than two 
gluons which combine into one. A remarkable property of 
of the EGLLA in the $C$-even sector is that reggeization leads to 
exact solutions of the integral equations for all odd numbers of 
gluons. That means that the amplitudes $D_{2l+1}$ can then be expressed 
in terms of the amplitudes $D_2, D_4, \dots, D_{2l}$. This was found 
explicitly for up to six gluons and argued to hold also for general $l$ 
in \cite{Bartels:1999aw}. 

Motivated by these observations in the $C$-even channel we 
now construct a four-gluon amplitude for the $C$-odd channel 
in the same spirit. More precisely, we will show that 
the following ansatz solves the integral equation (\ref{inteqf4}) 
for the full four-gluon state $F_4$: 
\begin{equation}
\label{f4ansatz}
\eqeps{glf41}{-3} = \frac{g}{2} \left[
\eqeps{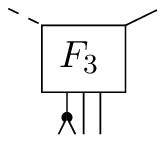}{-2}+
\eqeps{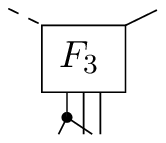}{-2}+
\eqeps{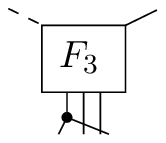}{-2}+
\eqeps{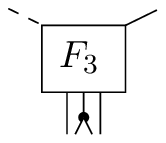}{-2}+
\eqeps{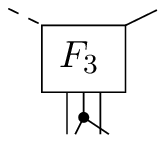}{-2}+
\eqeps{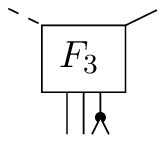}{-2}\right]\,.
\end{equation}
These $F_3$-diagrams will be called `splitting pairs' from now on.
The amplitudes $F_3$ again have three arguments, one of which is the sum of 
the two merged gluons. Again, a contraction of the color tensor of the
amplitude $F_3^{b_1 b_2 b_3}$ with the structure constant $f$ 
is implied. Writing out the ansatz explicitly, we have 
\begin{equation}
\begin{split}
\label{f4solutionexplicit}
  F_4^{a_1 a_2 a_3 a_4}=\frac{g}{2}
& \left[ f_{a_1 a_2 k} F_3^{k a_3 a_4}(12,3,4) + f_{a_1 a_3 k} F_3^{k a_2 a_4}(13,2,4)
+ f_{a_1 a_4 k} F_3^{k a_2 a_3}(14,2,3) \right.\\
+ & \left. f_{a_2 a_3 k} F_3^{a_1 k a_4}(1,23,4)
+ f_{a_2 a_4 k} F_3^{a_1 k a_3}(1,24,3) + f_{a_3 a_4 k} F_3^{a_1 a_2 k}(1,2,34)\right] \,.
\end{split}
\end{equation}
Note that in this ansatz the full amplitude $F_4$ is expressed in terms of 
$F_3$ in exactly the same way as in our result (\ref{f40ansatz}) the four-gluon 
impact factor $F_{(4;0)}$ was given 
in terms of the three-gluon impact factor $F_{(3;0)}$. 

The above ansatz already leads to a very simple cancellation 
in the four-gluon Odderon 
equation (\ref{inteqf4diag}). One expresses all amplitudes $F_4$ 
and the quark loop $F_{(4;0)}$ in terms of three-gluon diagrams.
Then one can use the BKP equation (\ref{inteqf3}) to evaluate the term 
$\omega F_4^{a_1 a_2 a_3 a_4}$ and obtains exactly the same expression for 
the quark loop as on the right hand side of the equation. Thus, one is left
with an equation in which only the full three-gluon amplitude $F_3$ appears. 

We now consider one particular splitting pair of gluons. 
When using the BKP equation to cancel the $\omega$-term
and the quark loop, one gets additional terms involving the three-gluon
amplitude $F_3$ in which the corrections due to the kernel or the $\beta$
function occur before the splitting of one gluon into two.
There are three virtual ($\beta$) corrections of this kind: two
where the $\beta$-term acts on a gluon that does not split and
one where it acts on the gluon that further down splits into two gluons.
In the four-gluon equation there are four virtual corrections. 
Two of these act on gluons coming directly from the amplitude $F_4$, 
i.\,e.\ that have not emerged from a splitting. These cancel
the corresponding two terms coming from the three-gluon equation.
The other two virtual corrections involve the two gluons that have
emerged from the gluon splitting. One then easily finds that these terms
together with the upper correction of the splitting gluon exactly
cancel the contribution of the $2\to 2$ kernel acting on the two 
gluons that have emerged from the splitting.

This means that all the virtual corrections and the term in which the splitting
and the kernel take place in the same two gluons have canceled. We have
considered only one particular pair of splitting gluons but the 
cancellation immediately extends to the sum of all splittings. 
Now one is left exactly with all the terms involving the $2 \to 3$ kernel
and the remaining $2 \to 2$ terms. 

At this stage one can collect all the diagrams in which 
one particular gluon, say the $i$th
outgoing gluon, passes the kernel and the splitting without being affected. 
Then one is left with four groups of identical expressions that cancel
separately due to an identity that was already discussed in the context
of reggeization in the Pomeron in \cite{Bartels:1999aw}, and in a 
somewhat different exposition also in \cite{Braun:1995hh}. 
We show this identity in pictorial language and omit the unaffected gluon:
\begin{eqnarray}
 \picb{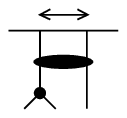} + \picb{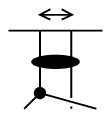} 
+ \picb{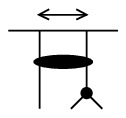} &=&
 \frac{2}{g}\,\picb{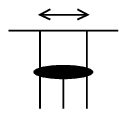} + \picb{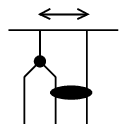}
+ \picb{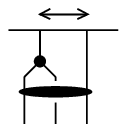}
\nonumber \\
&& + \picb{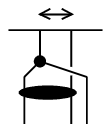}+ \picb{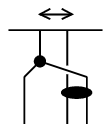}
+ \picb{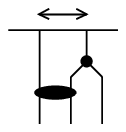}+ \picb{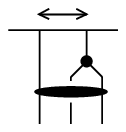}
\,.
\label{reggeizebilder}
\end{eqnarray}
The arrows on top of the diagrams symbolize the symmetry of the amplitude
under simultaneous exchange of momentum and color labels. In our case the
interactions mediated by the integral kernel get contracted with the three
gluon Odderon amplitude
which also exhibits this symmetry as we have explained earlier. 
That symmetry is the only property of the amplitude that is needed for the proof 
of this identity.
That completes our proof that the ansatz (\ref{f4ansatz}) for $F_4$ 
in fact solves the full integral equation (\ref{inteqf4}) for the four-gluon 
Odderon state in the EGLLA. 

Our result establishes that gluon reggeization takes place in the $C$-odd 
channel in the same way as in the $C$-even channel. 
At the same time, it shows that one cannot couple an actual four-gluon 
state to the $\gamma \to \eta_c$ impact factor in EGLLA. Instead, 
the impact factor couples only to a superposition of three-gluon Odderon 
states. 

An important property of our solution for $F_4$ is 
that it satisfies the same Ward-type identities which were found for 
the $n$-gluon amplitudes in the $C$-even sector in  \cite{Ewerz:2001fb}. 
These identities follow directly from (\ref{f4ansatz}) and 
from the fact that $F_3$ vanishes if one of the three gluons carries zero 
transverse momentum. More precisely, we find that $F_4$ vanishes 
if the first or last gluon momentum is zero, 
\begin{equation}
\left. 
F_4(\vc k_1,\vc k_2, \vc k_3,\vc k_4) \right|_{\vc k_1=0} = 
\left.
F_4(\vc k_1,\vc k_2, \vc k_3,\vc k_4) \right|_{\vc k_4=0} = 0
\,,
\end{equation}
whereas for the case that the inner gluon's momenta vanish 
the amplitude reduces to the lower amplitude $F_3$. The crucial 
point here is the specific behavior of the color tensors in this 
reduction. For the case $\vc k_2=0$ we find 
\begin{equation} 
\left.
F_4^{a_1 a_2 a_3 a_4}(\vc k_1,\vc k_2, \vc k_3,\vc k_4) 
\right|_{\vc k_2=0} =
g f_{a_1 a_2 k} F_3^{k a_3 a_4} (\vc k_1, \vc k_3,\vc k_4) 
\,,
\end{equation}
or alternatively
\begin{equation}
\left.
F_4^{a_1 a_2 a_3 a_4}(\vc k_1,\vc k_2, \vc k_3,\vc k_4) 
\right|_{\vc k_2=0} =
g f_{a_2 a _3 k} F_3^{a_1 k a_4} (\vc k_1, \vc k_3,\vc k_4) 
+  g f_{a_2 a _4 k} F_3^{a_1 a_3 k} (\vc k_1, \vc k_3,\vc k_4) 
\,.
\end{equation}
To obtain the last two identities we have made use of the 
overall color neutrality of the four gluons. 
Similarly, we find for the case $\vc k_3=0$ 
\begin{equation}
\left.
F_4^{a_1 a_2 a_3 a_4}(\vc k_1,\vc k_2, \vc k_3,\vc k_4) 
\right|_{\vc k_3=0} =
g f_{k a_3 a_4} F_3^{a_1 a_2 k} (\vc k_1, \vc k_2,\vc k_4) 
\,,
\end{equation}
which also equals 
\begin{equation} 
\left.
F_4^{a_1 a_2 a_3 a_4}(\vc k_1,\vc k_2, \vc k_3,\vc k_4) 
\right|_{\vc k_3=0} =
g f_{a_2 a _3 k} F_3^{a_1 k a_4} (\vc k_1, \vc k_2,\vc k_4) 
+  g f_{a_1 a _3 k} F_3^{k a_2 a_4} (\vc k_1, \vc k_2,\vc k_4) 
\,.
\end{equation}
These are exactly the same identities that also hold for the 
reggeizing part of the four-gluon amplitude $D_4$ in the 
$C$-even channel, that is for the part which does not contain 
the two-to-four gluon vertex. 
For a detailed discussion of the Ward-type identities and of their 
significance for 
finding an effective field theory of interacting reggeized gluons 
we refer the reader to \cite{Ewerz:2001fb}. 

Note that up to now we have obtained 
our results only for the BLV Odderon solution \cite{Bartels:1999yt}. 
This is because we have considered the 
integral equations of the EGLLA for the case of the 
$\gamma \to \eta_c$ impact factor, and the BLV solution is the 
only known Odderon solution which couples to that impact factor 
in leading order. 
In the next section we will show that an analogous solution for the 
four-gluon amplitude exists also for the Janik-Wosiek (JW) Odderon 
solution \cite{Janik:1998xj}. 

\section{Generalization to other Odderon solutions}
\label{sec:jw}

The BLV solution that we have discussed so far is not the only possible 
solution of the BKP equation for the $C$-odd three-gluon state. 
The other class of solutions has the characteristic property that 
it vanishes if two of the transverse gluon coordinates coincide. 
That class comprises the Janik-Wosiek Odderon solution 
\cite{Janik:1998xj}, and we will refer to that class generically 
as the JW Odderon. In order to study that class of solutions 
in the EGLLA we have to make a different choice for the 
inhomogeneous term of the integral equations that we have 
discussed in section \ref{sec:eveq}. In order to couple the JW 
Odderon to an impact factor in leading order we should consider 
an impact factor in which the three gluons are coupled to 
three different partons, i.\,e.\ we need three different 
points in transverse space to couple the gluons to. A suitable choice 
is a baryonic impact factor. The use of perturbation theory is in general 
questionable for light baryons. For the purpose of the 
present paper, however, we can think of a heavy baryon or of a 
large momentum transfer $\sqrt{t}$ in the case of a light baryon. 
Note that the integral equations of section \ref{sec:eveq} remain 
fully valid after replacing the $\gamma \to \eta_c$ impact 
factors $F_{(n;0)}$ by suitable baryonic impact factors. 

In contrast to the photon and the $\eta_c$ meson baryons are 
not eigenstates of $C$-parity. Baryonic impact factors therefore 
contain both $C$-even and $C$-odd contributions. A full discussion 
of the baryonic impact factor with arbitrary numbers of gluons 
is beyond the scope of the present paper. Instead, we will restrict 
ourselves to the impact factor relevant for the JW solution and 
for up to four gluons. We hence consider only diagrams in which 
at least one gluon is attached to each of the three quarks in the baryon, 
since all other diagrams would vanish when folded with 
the JW solution. The diagrams we have to consider for three 
gluons are hence of the type (to be specific here we choose a 
$p \to  p$ impact factor) 
\begin{equation}
\label{diagproton3}
E^{abc}(\vc k_a, \vc k_b, \vc k_c) = \,\,\,\,
\eqepsres{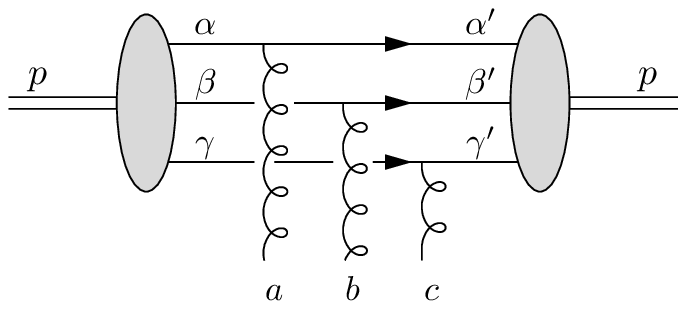}{6}{5.8cm} \,.
\end{equation}
Incidentally, this diagram contributes only to the $C$-odd channel. 
This is most easily seen from its color factor, 
\begin{equation}
\label{colordiagproton3}
\varepsilon_{\alpha \beta \gamma} \,
t^a_{\alpha' \alpha} t^b_{\beta' \beta} t^c_{\gamma' \gamma} \,
\varepsilon_{\alpha' \beta' \gamma'} = \frac{1}{2} d_{abc}
\,,
\end{equation}
which contains only the symmetric color configuration of the 
three gluons. The full relevant impact factor $B^{p \to p\, JW}_{(3;0)}$ 
is obtained by adding all diagrams of this type. 
(For a baryon with three different 
constituent quark flavors we have six diagrams, otherwise 
it can be less with suitable symmetry factors for equal quark 
flavors.) After this summation the impact factor is symmetric 
in the color labels and in the momenta of the three gluons separately. 
Since our diagrams contribute only to the $C$-odd 
channel it is clear that the corresponding diagrams for the 
$\bar{p} \to \bar{p}$ impact factor have the opposite sign, but 
are otherwise identical. This is again due to the sign change 
of the vertices when replacing all quarks by antiquarks. 
Therefore the $C$-odd impact factor for the JW Odderon 
solution is identical to the 
$p \to p$ impact factor in this case, 
\begin{equation}
B^{{\cal O}\, JW}_{(3;0)} = \frac{1}{2} \left[
B^{p \to p\, JW}_{(3;0)} - 
B^{\bar{p} \to \bar{p}\, JW}_{(3;0)}
\right]
= B^{p \to p\, JW}_{(3;0)} \,.
\end{equation}

As in the case of the $\gamma \to \eta_c$ impact factor, we will not 
need the specific dependence of the diagram $E$ in (\ref{diagproton3}) 
on the gluon momenta, for phenomenological models we refer 
the reader to \cite{Fukugita:1979fe} and \cite{Levin:gg} 
(see also \cite{Dosch:2002ai}). Instead, we will again make use 
of the reduction mechanism explained in section \ref{sec:mechanism} 
and express the relevant diagrams with four gluons in terms of 
three-gluon diagrams of the type (\ref{diagproton3}). Note that 
the same cuts are applied to the amplitudes 
as before such that the quarks between the gluons are set on-shell. 
Therefore all conditions for the reduction mechanism are fulfilled and 
we can apply it to each pair of gluons coupled to the same quark line. 

Let us now consider the baryonic impact factor with four gluons. 
Given the two classes of Odderon solutions with three gluons it is 
natural to expect that there will also be two distinct classes of solutions 
in the case of four gluons. Here we concentrate only on the 
JW type solution, and in analogy to the three-gluon case we 
consider only diagrams contributing to the impact factor in which each 
quark has at least one gluon coupled to it. In the case of four gluons 
the resulting diagrams still contain both $C$-even and $C$-odd 
contributions. We therefore need to make a projection onto the 
$C$-odd amplitude by taking one half of the difference between the 
baryonic impact factor and its anti-baryonic analog. 

Obviously, in all possible diagrams there is exactly one quark to 
which two gluons are attached. We hence have a sum 
over all six possible pairs of gluons. As an example we consider 
the first two gluons being coupled to the same quark. A typical 
diagram is 
\begin{equation}
\label{diagproton4_1}
E_4^{p\to p} = \,\,\,\,
\eqepsres{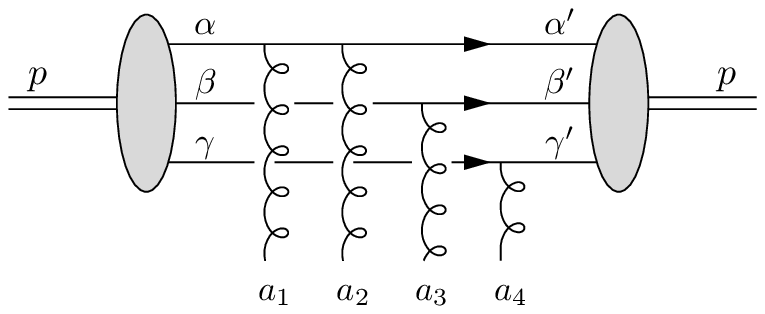}{6}{6.48cm} \,,
\end{equation}
and we have to sum over all possible permutations of the quark 
lines exactly as in the three-gluon impact factor. 
The color part of that diagram is 
\begin{equation}
\varepsilon_{\alpha \beta \gamma} \,
(t^{a_2} t^{a_1})_{\alpha' \alpha} 
t^{a_3}_{\beta' \beta} t^{a_4}_{\gamma' \gamma} \,
\varepsilon_{\alpha' \beta' \gamma'} 
\,, 
\end{equation}
since the three quarks are in an antisymmetric color state in 
the baryon. 

In the case of the impact factor with an anti-baryon, the possible 
diagrams are obtained in the same way. The diagram corresponding 
to the one above is 
\begin{equation}
E_4^{\bar{p} \to \bar{p}} =\,\,\,\,
\eqepsres{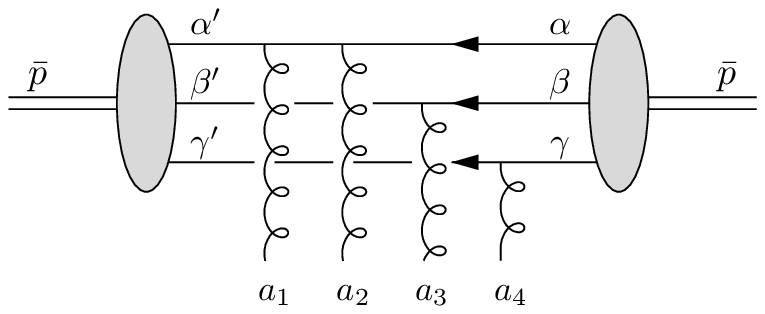}{6}{6.48cm} \,.
\end{equation}
For later convenience we have already relabeled the summation 
indices for the colors in the incoming and outgoing baryon. 
The color factor for this diagram is 
\begin{equation}
\varepsilon_{\alpha' \beta' \gamma'} \,
(t^{a_1} t^{a_2})_{\alpha' \alpha} 
t^{a_3}_{\beta' \beta} t^{a_4}_{\gamma' \gamma} \,
\varepsilon_{\alpha \beta \gamma} 
\,.
\end{equation}

The momentum parts of both diagrams are identical, in particular 
there is no sign change (in contrast to the three-gluon diagram, see 
above) because we have an even number of gluons here. We can 
now apply the reduction formula (\ref{diagreduction}) and find 
that the momentum part of both diagrams is $ig$ times the 
momentum part of the three-gluon diagram $E$ of (\ref{diagproton3}) 
with the three momentum arguments $(\vc k_1 + \vc k_2)$, $\vc k_3$, and 
$\vc k_4$. The difference of the color factors of the two diagrams, 
on the other hand, is 
\begin{equation}
\frac{1}{2} \, \varepsilon_{\alpha \beta \gamma} \,
\left( t^{a_2} t^{a_1} - t^{a_1} t^{a_2} \right)_{\alpha' \alpha} 
t^{a_3}_{\beta' \beta} t^{a_4}_{\gamma' \gamma} \,
\varepsilon_{\alpha' \beta' \gamma'} 
= - \frac{1}{2} i f_{a_1 a_2 k} \,
\varepsilon_{\alpha \beta \gamma} \,
t^k_{\alpha' \alpha} t^{a_3}_{\beta' \beta} t^{a_4}_{\gamma' \gamma} \,
\varepsilon_{\alpha' \beta' \gamma'} \,,
\end{equation}
where we have expressed that difference in terms of the color factor 
(\ref{colordiagproton3}) of the three-gluon diagram $E$. 

Hence the total contribution of our two diagrams to the $C$-odd channel 
is 
\begin{eqnarray}
&&\frac{1}{2} \,
\left[ \left(E_4^{p \to p}\right)^{a_1a_2a_3a_4} 
(\vc k_1, \vc k_2, \vc k_3, \vc k_4) 
- \left(E_4^{\bar{p} \to \bar{p}}\right)^{a_1a_2a_3a_4} 
(\vc k_1, \vc k_2, \vc k_3, \vc k_4) 
\right] = 
\nonumber
\\
&& \hspace*{1cm}
= \frac{1}{2} \,
g f_{a_1 a_2 k} E^{k a_3 a_4}(\vc k_1 + \vc k_2, \vc k_3, \vc k_4) 
\,.
\end{eqnarray}
The same mechanism obviously holds for the other permutations of the 
quark (respectively antiquark) lines, and the last formula then naturally 
extends to the sum of those diagrams. It remains to sum over all 
pairs of gluons, and for each pair of gluons we obtain an analogous 
expression. Denoting the full impact factor obtained in this way 
by $B_{(4;0)}^{{\cal O}\,JW}$ we can write it in terms of the 
corresponding three-gluon impact factor $B_{(3;0)}^{{\cal O}\,JW}$ as 
\begin{equation}\label{B40ansatzsum}
B_{(4;0)}^{{\cal O}\,JW} = 
\frac{g}{2}\sum_{i,j \in \{1,\ldots,4\} \atop i < j} 
\left[\eqeps{3to4-12}{-2} \right]
\star \left(B^{{\cal O}\,JW}_{(3;0)}\right)^{b_1 b_2 b_3} (i j) \,,
\end{equation}
where we have again used the notation introduced 
in section \ref{sec:loopfour}. 
We recall that the superscript $JW$ refers to the fact that we have 
restricted our analysis to diagrams in which the gluons are 
coupled to all three quarks, that is to three distinct points 
in transverse space, hence giving rise to the JW Odderon solution. 
Note that this formula is identical to the way (\ref{F40ansatzsum}) 
in which the $\gamma \to \eta_c$ impact factor with four 
gluons was expressed in terms of the one with three gluons. 
Clearly, reggeization takes place in the same way in both impact 
factors. 

Let us now proceed to the full three- and four-gluon amplitudes arising 
from the part of the baryonic impact factor relevant to the JW Odderon. 
The coupled integral equations for these amplitudes 
$B_3^{{\cal O}\,JW}$ and $B_4^{{\cal O}\,JW}$ are completely 
analogous to those for $F_3$ and $F_4$, see (\ref{inteqf3}) and 
(\ref{inteqf4}), with the $\gamma \to \eta_c$ impact factors replaced by the 
impact factors $B_{(3;0)}^{{\cal O}\,JW}$ and $B_{(4;0)}^{{\cal O}\,JW}$ 
discussed above. Since the latter two impact factors are related to 
each other in exactly the same way as the three- and four-gluon 
$\gamma \to \eta_c$ impact factors we can immediately apply the 
result of section \ref{sec:solution}. Hence we find a solution for the 
full four-gluon amplitude $B_4^{{\cal O}\,JW}$ as 
a superposition of three-gluon amplitudes 
$B_3^{{\cal O}\,JW}$ in exactly the way given in (\ref{f4ansatz}) or 
in (\ref{f4solutionexplicit}). We conclude that also in the JW class of 
Odderon solutions an actual four-gluon state does not contribute to 
the amplitude. Instead, due to reggeization the four-gluon amplitude 
has the analytic properties induced by the 
corresponding three-gluon amplitude. Again, reggeization takes place 
in the same way as in the $C$-even sector. Also here the Ward-type 
identities of \cite{Ewerz:2001fb} are fulfilled. 

Two more remarks are in order concerning the baryonic impact 
factor and the way in which we have computed it. 
In the present section we have obtained the $C$-odd contribution 
to the impact factors 
as the difference of the baryonic impact factor and its $C$-conjugate, 
that is by explicitly projecting onto the $C$-odd channel. 
That method was also used (in position space) in \cite{Kovchegov:2003dm} 
where the BLV Odderon solution was found in the dipole picture. 
Implicitly, we have done something similar in the calculation of the 
quark loop in the $\gamma \to \eta_c$ impact factor in section 
\ref{sec:impfac} above. There it was the $\gamma^5$ matrix in the loop 
which caused the relative signs between two diagrams related to each 
other by the exchange of quark and antiquark line (corresponding 
to a $C$-parity transformation of the quark-antiquark intermediate states). 
We found it more convenient to stay as close as possible to previous 
calculations of the $\gamma \to \eta_c$ impact factor with three gluons. 
Baryonic impact factors always contain both $C$-even and $C$-odd 
contributions and therefore in this case the explicit projection is needed, 
in contrast to the $\gamma \to \eta_c$ impact factor which 
couples only to $C$-odd states of gluons. 

The calculation of the full baryonic impact factors for arbitrary numbers 
of gluons is clearly feasible with the methods presented here, both for 
$C$-even and for $C$-odd states of gluons. The calculation of those 
impact factors would be important for a better understanding of 
reggeization of the gluon in the high energy limit. We expect that 
taking into account also contributions in which not all quark lines have 
gluons attached one would obtain couplings to both classes of Odderon 
solution, that is JW and BLV solutions. It should then be possible 
to obtain full solutions of the integral equations for four gluons as 
superpositions of full JW and BLV solutions in analogy to the solutions 
found here. That would provide further evidence for the universality 
of gluon reggeization at high energies. 

\section{Summary and outlook}
\label{sec:summary}

In this paper we have made the first step in a systematic investigation 
of $C$-odd exchanges in the color glass condensate. We have used the 
EGLLA, that is the perturbative 
approach based on the resummation of logarithms of the energy 
where we allow the number of gluons to fluctuate during the $t$-channel 
evolution. The lowest amplitude is the well-known Odderon consisting 
of three reggeized gluons. In the present paper we have discussed the 
four-gluon amplitude. We have computed the coupling of four gluons 
to the $\gamma \to \eta_c$ impact factor and to a baryonic impact factor. 
Using these results we have found exact solutions for 
the corresponding four-gluon amplitudes as superpositions of 
three-gluon amplitudes. In each term of that superposition 
a pair of gluons merges and behaves like a single gluon in the three-gluon 
amplitude. This reggeization happens in exactly the same 
way as in the amplitudes in the $C$-even sector, strengthening 
the evidence for the universality of this mechanism. Our result 
implies that there is no direct coupling of an actual four-gluon state 
to the relevant impact factors at the leading logarithmic level. 
Instead, the analytic properties of the $C$-odd four-gluon amplitude 
in the EGLLA are fully determined by those of the three-gluon amplitude. 

It has been found that in the $C$-even channel the EGLLA is dominated 
by exchanges with even numbers of gluons. More precisely, all amplitudes 
with odd numbers of gluons in the $t$-channel reggeize and become 
superpositions of amplitudes with even numbers of gluons. Hence the 
energy dependence is fully given by the spectrum of the $n$-gluon states 
with even $n$. Correspondingly, the $C$-even amplitudes can be cast 
into the form of an effective field theory in which only 
even $n$-gluon states occur which couple to each other via effective 
transition vertices $V_{2 \to 2l}$ \cite{Bartels:1999aw}. 
Moreover, it has been observed that the $n$-gluon states as well as the 
effective transition vertices are conformally invariant in two-dimensional 
impact parameter space. The obvious goal is to identify that effective 
conformal field theory of reggeized gluons. 
Note that this structure of an effective field theory emerges only if one 
takes into account subleading terms in $N_c$ in the $n$-gluon amplitudes. 
The leading contributions in the expansion in $1/N_c$ reggeize completely 
when coupled to a photon-impact factor, see for example 
\cite{Braun:1997gm} and \cite{Bartels:1999aw}. 
In order to study the interesting possibility of an effective conformal field 
theory of high energy QCD it is therefore crucial to go beyond the 
large-$N_c$ approximation. 

The picture of an effective field theory of reggeized gluons in the $C$-even 
channel suggests that a similar structure should also emerge in the 
$C$-odd channel. Our results indicate that in the $C$-odd channel only 
states with odd numbers of gluons occur, and that here the amplitudes 
with even number of gluons like $F_4$ are superpositions of amplitudes 
with odd numbers of gluons. In that sense the study of the $C$-odd 
amplitudes is complementary to the investigation of the $C$-even channel. 
We expect that in the $C$-odd five-gluon 
amplitude a new effective three-to-five gluon vertex should appear. 
In the present paper we have shown the main condition for this to 
be fulfilled, 
namely the reggeization of the gluon in the $C$-odd amplitudes. 
Recall that the transition kernels $K_{2 \to m}$ in the integral equations 
of the EGLLA always start from two gluons. 
We therefore expect that the three-to-five transition proceeds 
via independent two-to-four gluon transitions in each gluon 
pair. Since in the three-gluon amplitude each pair of gluons is in 
a symmetric color octet state that would give us something like an 
effective two-to-four gluon vertex in the symmetric octet. This in turn 
would be an element of the effective theory which should play an 
important role also in the $C$-even channel, for example in the 
transition from the four- to the six-gluon state. The study of the 
$C$-odd five-gluon amplitude would therefore contribute to 
a better understanding of the general picture of the effective 
field theory of the color glass condensate. 

The analysis of the integral equation for the five-gluon amplitude 
will clearly require the knowledge of the corresponding impact 
factor. In the present paper we have given an explicit formula for it 
in the case of the $\gamma \to \eta_c$ impact factor. 
In that calculation we have found different representations 
which are equivalent. In the approach to the integral equations 
developed in \cite{Bartels:1999aw}, however, the impact factor 
is promoted to an ansatz for the reggeizing part of the full 
amplitude. In that step, the different representations for the 
impact factor will in general yield different results, posing a 
potential problem for the further analysis of the integral 
equations. The origin of the different representations was 
the fact that in the $\gamma \to \eta_c$ impact factor the gluons 
come naturally in two groups according to whether they are 
coupled to the quark or the antiquark, but on the other hand 
we need to split them into three groups for having three momentum 
arguments of the Odderon amplitude. This problem can be 
avoided by using baryonic impact factors instead. 
Therefore we consider it an important future project to 
generalize our analysis of the baryonic impact factor to 
arbitrary numbers of gluons. This would give us more information 
on the reggeization of the gluon and might be interesting also in 
the $C$-even channel. 

The knowledge of the three-to-five gluon transition in the 
effective field theory for the color glass condensate would 
be very interesting from several points of view. It would 
for example allow one to compute the splitting of an 
Odderon into an Odderon plus a Pomeron. From the theoretical 
perspective, one could compare it with 
the known splitting vertex of the Pomeron into two Odderons 
and hence study the properties of the effective vertices under crossing. 
Also from a phenomenological perspective it would be useful 
to compute that vertex. Clearly, it would help us in understanding 
the interplay of Pomeron and Odderon exchanges in high energy 
reactions. Further, it would also make it possible to compute the 
effect of Pomeron loops on the Odderon intercept, and to study 
saturation effects for the Odderon in the large-$N_c$ limit. 

\section*{Acknowledgements}

We would like to thank O.\ Nachtmann and S.\ Wallon for helpful discussions. 
C.\,E.\ was supported by the Bundesministerium f\"ur 
Bildung und Forschung, projects HD 05HT1VHA/0 
and HD 05HT4VHA/0, and by a Feodor Lynen fellowship of the 
Alexander von Humboldt Foundation. 

\begin{appendix}

\boldmath
\section{The Pomeron quark loop with $n$ gluons}
\unboldmath
\label{app:pomloop}

In this appendix we want to consider the $\gamma^* \to \gamma^*$ 
impact factor with $n$ gluons attached. Due to the quantum numbers 
of the photon the $n$-gluon system is in a $C$-even state and hence 
contributes to the Pomeron channel. This particular impact factor was 
already considered in \cite{Bartels:1999aw} for up to six gluons in the 
context of the EGLLA. In \cite{Braun:1995hh} the same impact factor 
was studied in the simpler case of the large-$N_c$ limit. 
The considerations in the present paper make 
it now straightforward to obtain the general form of that impact factor 
for arbitrary gluon numbers. 

There are some differences to the $\gamma \to \eta_c$ impact 
factor. Firstly, the lowest Pomeron state is built of only two gluons and the
color tensor is $\delta^{a_1 a_2}$ for this state. The corresponding two-gluon
Pomeron quark loop is called $D^{a_1 a_2}_{(2;0)}$ and the momentum part
$D_{(2;0)}$ consists of only two diagrams. (Similar to the Odderon 
case described above, there are originally four diagrams that can be reduced 
to two when taking into account the symmetry under quark-antiquark 
exchange.) 
Another difference concerns the
color tensors. Our analysis of the $\gamma \to \eta_c$ impact factor 
made use of the fact that there 
is a $\gamma^5$ matrix at the meson vertex and therefore a mass term must 
be chosen in one quark propagator to give a non-vanishing $\gamma$-trace. In 
the Pomeron case this is no longer true, as there is a $\gamma^\mu$ 
instead of the $\gamma^5$ matrix and all propagators therefore lend their 
momentum parts in order to give a non-vanishing trace (in the limit that the 
quark mass is negligible compared to the typical longitudinal quark momenta). 
This implies that now 
states with an odd number of attached gluons come with $f$ type tensors,
whereas states with an even number of gluons come with $d$ type tensors. 

The main difference becomes clear when considering the state with three 
$t$-channel gluons. For the Odderon we 
already found two different color tensors when we added one gluon to 
the minimal number of gluons, that is in the four-gluon
quark loop. In Pomeron case, on the contrary, 
the three-gluon Pomeron quark loop has only 
one color tensor $f^{a_1 a_2 a_3}$. In complete analogy to (\ref{f40ansatz}) 
the three-gluon Pomeron quark loop can be constructed by considering the three 
possible combinations of splitting gluons. An $f$ tensor is again contracted
with the color tensor of the amplitude:
\begin{align}
  D_{(3;0)}^{a_1 a_2 a_3}(\vc k_1, \vc k_2, \vc k_3)  = \frac{g}{2} 
  & \left [ f_{a_1 a_2 b_1} \delta^{a_3 b_2} D_{(2;0)}^{b_1 b_2}(12,3) 
\nonumber \right. \\
 & + \left. f_{a_1 a_3 b_1} \delta^{a_2 b_2} D_{(2;0)}^{b_1 b_2}(13,2)+
  f_{a_2 a_3 b_2} \delta^{a_1 b_1} D_{(2;0)}^{b_1 b_2}(1,23)\right ] \,.
\end{align}
Again, the arguments denote the indices of the momenta that enter as a sum
in the two gluon amplitude. As $D_{(2;0)}$ is symmetric under exchange of its
momentum arguments, we do not need to care about the order of its 
arguments. 
Similarly to the Odderon case, the above expression 
can be written in terms of the momentum part 
of the lowest impact factor only:
\begin{equation}
  D_{(3;0)}^{a_1a_2a_3}(\vc k_1,\vc k_2,\vc k_3) 
= \frac{g}{2} f_{a_1a_2a_3} \,
[ D_{(2;0)}(12,3) - D_{(2;0)}(13,2) + D_{(2;0)}(1,23) ]  \,. 
\label{d30}
\end{equation}
In the case of four gluons the amplitude contains two different color 
structures. The corresponding momentum structures are again abbreviated
as in the Odderon case in (\ref{f40}), 
\begin{eqnarray}
D_{(4;0)}^{a_1a_2a_3a_4}(\vc k_1,\vc k_2,\vc k_3,\vc k_4)
 &=& \! - g^2 d^{a_1a_2a_3a_4} \, [ D_{(2;0)}(123,4) + D_{(2;0)}(1,234) 
                                  - D_{(2;0)}(14,23) ] 
   \nonumber \\
 & & \! - g^2 d^{a_2a_1a_3a_4}  \, [ D_{(2;0)}(134,2) + D_{(2;0)}(124,3)
                                 - D_{(2;0)}(12,34) \nonumber \\
 && \hspace{2.4cm} - D_{(2;0)}(13,24) ]
\nonumber \\
&\equiv& d^{a_1a_2a_3a_4} D_{(4;0)}^{(1)}
 + d^{a_2a_1a_3a_4}  D_{(4;0)}^{(2)} \,.
\label{d40}
\end{eqnarray}
The $n$-gluon quark loop for the Pomeron ($n\geq 4$) 
is then constructed in complete analogy 
to the Odderon quark loop in (\ref{ngluoncomplete})
as a superposition of these terms $D_{(4;0)}^{(1)}$
and $D_{(4;0)}^{(2)}$. Clearly, the tensor $-i f$ now appears in the odd
and the $d$ tensors in the even gluon-number amplitudes.
These results are in complete agreement with the ones found in 
\cite{Bartels:1999aw} for up to $n=6$. 

\end{appendix}

\end{document}